\documentclass{optica-article}

\journal{opticajournal} 

\articletype{Research Article}
\usepackage{graphicx}
\usepackage{lineno}

\begin{document}

\title{Pulse-to-pulse Power Fluctuations in Ultra-Fast Fiber Amplifiers with Gain-Managed Nonlinearity}

\author{Maksim Kozlov,\authormark{1,*} Colin J. McKinstrie,\authormark{2} Nitzan Haviv\authormark{3} and Pavel Sidorenko\authormark{3}}

\address
{\authormark{1}Center for Preparatory Studies, Nazarbayev University, 010000, Astana, Republic of Kazakhstan \\
\authormark{2}Independent Photonics Consultant, Manalapan, NJ 07726, USA \\
\authormark{3} Department of Electrical and Computer Engineering and Solid-State Institute, Technion– Israel Institute of Technology, 32000 Haifa, Israel }

\email{\authormark{*}maksim.kozlov@nu.edu.kz} 

\begin{abstract*} 
We present a theoretical framework for predicting pulse-to-pulse power fluctuations caused by pump and seed power noise and by amplified spontaneous emission in ultrafast fiber amplifiers operating in the gain-managed nonlinear regime. The numerical model evaluates statistically averaged quantities directly, without Monte Carlo simulations with stochastic sources, and reduces to a semi-analytical monochromatic model in the limit of vanishing signal bandwidth and negligible ASE. We show that nonlinear spectral broadening of the signal and of its perturbations, combined with the frequency-dependent gain, significantly alters the technical relative intensity noise in both the high-power, high-repetition-rate and the low-power, low-repetition-rate regimes, and that in the latter regime the contributions related to amplified spontaneous emission become comparable to those induced by pump and seed fluctuations. 

\end{abstract*}

\section {Introduction}
Ultrafast fiber lasers have become attractive sources for industrial, biomedical, and scientific applications owing to their compactness, alignment-free operation, excellent beam quality, and low manufacturing cost \cite{Zhao, Ferm1, Ferm2, Chang}. Their performance, however, is fundamentally limited by nonlinear effects that arise when high-peak-power pulses propagate through the small core of an optical fiber over extended distances \cite{Perry}.

The conventional approach for suppressing nonlinear distortions is chirped-pulse amplification (CPA) \cite{Morou}, in which pulses are temporally stretched before amplification and recompressed afterward. Although highly successful, CPA is ultimately constrained by gain narrowing and residual dispersion mismatch, typically limiting compressed pulse durations to approximately 200 fs \cite{Zhao}. Generating substantially shorter pulses requires nonlinear compression \cite{Shank}, which increases system complexity and reduces efficiency. Alternative techniques, including direct amplification \cite{Zaout}, pre-chirp management \cite{Liu1, Chen}, and self-similar pulse evolution \cite{And, Krug, Ferm}, have enabled sub-100-fs pulse generation \cite{Liu1, Liu2, Song}, but often at the expense of increased design complexity and reduced operating flexibility.

A promising alternative is the recently demonstrated gain-managed nonlinearity (GMN) regime \cite{Sidor, Sidor2}. GMN regime exploits the interplay between self-phase modulation and a longitudinally evolving asymmetric gain spectrum to form a nonlinear attractor, enabling stable amplification over a broad range of seed conditions. This regime combines high robustness with compatibility with fully integrated all-fiber architectures \cite{Xia}. In contrast to self-similar pulse evolution, where the pulse bandwidth must remain well below the gain bandwidth to ensure a nearly frequency-independent gain, the GMN regime exploits nonlinear spectral broadening (NSB) to increase the signal bandwidth until it becomes comparable to the gain bandwidth. This enables the generation of nearly transform-limited pulses with durations as short as 30 fs \cite{Sidor}. The gain of amplifiers in GMN regime can exceed 30 dB, allowing efficient amplification of low-energy pulses at low repetition rates \cite{Stol}, while its weak sensitivity to seed-pulse characteristics makes it well suited for amplifying commercially available frequency-comb lasers \cite{Tom}. Furthermore, the recent realization of erbium-doped GMN amplifiers extends these advantages to the telecommunications wavelength band \cite{Mik, Yang2026}. 

Although the GMN regime can be realized experimentally using relatively simple fiber-amplifier architectures, its complete theoretical description remains an open challenge. While intensity-noise transfer functions have been studied in continuous-wave and chirped-pulse Yb-doped fiber amplifiers within a monochromatic rate-equation framework \cite{Zhao2017, Gierschke2019, Han2021}, the noise characteristics of broadband ultrafast GMN amplifiers, including pulse-to-pulse power fluctuations, have not yet been systematically investigated. A detailed understanding of the mechanisms governing these fluctuations is essential for minimizing their impact in applications requiring ultralow intensity noise \cite{Kim2016}, such as optical frequency-comb amplification \cite{Did, Roh, Bra}, laser systems for gravitational-wave detection \cite{bode, Cah, Kwee, Buik}, and optical metrology \cite{Udem}. Notably, two recent independent experimental studies have demonstrated that nonlinear amplification in the GMN regime can suppress relative intensity noise below the input seed level in both Yb-doped ($1.03 \rm \mu m$) \cite{Nitzan} and Er-doped ($1.55 \rm \mu m$) \cite{Yang2026} fiber amplifiers, further highlighting the need for a comprehensive theoretical treatment.

In this work, we develop a rigorous theoretical framework for predicting pulse-to-pulse power fluctuations in gain-managed fiber amplifiers arising from the dominant noise sources: stochastic fluctuations of the input pump and seed powers, which are transferred to the output signal and to the amplified spontaneous emission (ASE), the signal–ASE and ASE–ASE beat noise, and shot noise. We revisit a semi-analytical model based on a monochromatic approximation of the signal field and quantify the impact of NSB, an inherent feature of GMN amplifiers, by comparing its predictions with those of a full numerical model. The latter solves the coupled propagation equations, namely a nonlinear Schrödinger equation for the signal field, transport equations for the forward- and backward-propagating ASE, and steady-state rate equations for the upper-level population, together with the equations governing the linear perturbations of these quantities. Importantly, the numerical approach evaluates statistically averaged quantities directly, thereby avoiding computationally intensive Monte Carlo simulations with stochastic source terms \cite{Paschotta2004}. As a result, we obtain the relative intensity noise (RIN) of the output signal associated with seed and pump power fluctuations (technical RIN), with the signal–ASE and ASE–ASE beat noise, and with shot noise.

We consider three distinct operating regimes:
(i) a linear regime, in which NSB is negligible and the numerical model reproduces the monochromatic approximation with high accuracy;
(ii) a nonlinear regime corresponding to high-power, high-repetition-rate operation; and
(iii) a nonlinear regime in low-power amplifiers, where nonlinear effects become significant due to reduced repetition rate. 

It is demonstrated that in the GMN regime the interplay between the NSB of the signal and of its perturbations and the frequency-dependent gain substantially modifies the technical RIN, both at high power and high repetition rate and at low power and low repetition rate. In the latter case ($\sim 0.5$ W, $\sim 0.25$ MHz) the ASE-related noise sources contribute to the RIN at a level comparable to that of pump and seed fluctuations. We also derive, to the best of our knowledge for the first time, analytical expressions for the signal- and pump-to-ASE noise transfer functions and refine the expressions for the signal–ASE and ASE–ASE beat-noise intensities.

We anticipate that these results provide a framework for understanding and mitigating noise in aforementioned applications of GMN fiber amplifiers.

\section{Exact Numerical Model}

The propagation of optical pulses in fibers with gain-managed nonlinearity, under the slowly varying envelope approximation, can be described by the generalized nonlinear Schrödinger equation (GNLSE) \cite{Agraval, Lindberg, DuttaGupta2022}:
\begin{equation}\label{GNLSE}
\psi_z= -i \beta_2 \psi_{tt}/2 + i \gamma |\psi|^2 \psi + A(t,z)/2,
\end{equation}
where $\psi=\psi(t,z)$ is the complex envelope of the electric field in the co-moving frame ( $t$ is fast retarded intra-pulse time), normalized so that $|\psi(t,z)|^2$ represents the optical power in watts. Parameters $\gamma$ and $\beta_2$ denote the nonlinear coefficient and the second-order group-velocity dispersion coefficient, respectively. The amplification function $A(t,z)$ is given by
\begin{equation} \label{AF}
A(t,z)=\mathcal{F}^{-1} \{ g \left( \omega, z \right) \tilde{\psi}  \left( \omega, z \right) \},
\end{equation}
where $g \left( \omega, z \right)$ is the position and frequency-dependent gain and the Fourier transform of the complex electric field given by $\tilde{\psi}\left( \omega,z \right)=\mathcal{F} \{  \psi(t,z) \}=\int e^{i \omega t} \psi(t,z) dt/\sqrt{2 \pi}$. The frequency dependence of gain must be taken into account in GMN systems, where the pulse bandwidth is limited by the gain bandwidth:
\begin{equation} \label{Gain}
g \left( \omega \right) \approx \Gamma_s  \left[ \sigma_e \left( \omega \right) n_2- \sigma_a \left( \omega \right) n_1 \right],
\end{equation}
where background losses were neglected, $\Gamma_s$ is the signal overlap with the doped core region of the fiber, $\sigma_e \left( \omega \right) $ and $\sigma_a \left( \omega \right)$ are frequency-dependent emission and absorption cross-sections, and $n_1$ and $n_2$ are effective lower- and upper-energy-level density populations, respectively. In the following, the upper-state population density will be referred to as the inversion. Under typical experimental conditions, the upper-level lifetime far exceeds the time between consecutive pulses, allowing for the steady-state approximation of the inversion \cite{Lindberg}:
\begin{equation} \label{Rateq2}
    \frac {\partial n_2  }{\partial t}=n_{\rm tot} R_a -n_2 \left( R_a+R_e+1/\tau_{ul}\right),
\end{equation}
where $\tau_{ul}$ is the upper-level lifetime and $n_{\rm tot}$ is the total doping concentration. Absorption ($R_a$) and emission ($R_e$) transition rates are comprised of transition rates due to signal, pump, forward and backward amplified spontaneous emission (ASE) 
\begin{equation} \label{total_rate}
R_{e(a)}=R^s_{e(a)}+R^p_{e(a)}+R^{f\ ase}_{e(a)}+R^{b\ ase}_{e(a)} 
\end{equation}
with 
\begin{align}
    & R^s_{e(a)}=\frac{\Gamma_s}{T_{\rm rep} \hbar A_d} \int \frac {\sigma_{e(a)} \left( \omega \right)|\tilde{\psi}\left( \omega \right)|^2} {\omega_s+\omega} d\omega \label{Rates} \\ 
    & R^p_{e(a)}=\frac{\Gamma_p \sigma_{e(a)} \left( \omega_p \right) P_p}{\hbar A_d \omega_p} \label{Ratep} \\  
    & R^{f(b)\ ase}_{e(a)}=\frac{\Gamma_s}{\hbar A_d} \int \frac {\sigma_{e(a)} \left( \omega \right)\rho^{f(b)} \left( \omega \right) } {\omega_s+\omega} d\omega \label{Ratea} ,
\end{align}
where $\omega_s$ is the carrier frequency of the signal, $T_{\rm rep}$ refers to the time interval between consecutive pulses in the pulse train (repetition time), $\Gamma_p$ is the pump overlap with the doped core region of the fiber, $\hbar$ is reduced Planck's constant, $A_d=\pi (d_{\rm core}/2)^2$ is dopant area, $P_p$ and $\omega_p$ are pump power and frequency, respectively and $\rho^{f(b)}$ is the power spectral density of the forward (backward) propagating ASE ($|\tilde{\psi}\left( \omega \right)|^2/T_{\rm rep}$ is the power spectral density of signal). Pump power satisfies a simple differential equation:
\begin{equation} \label{PumpPower}
d_z {P_p}= g_p P_p.
\end{equation}
Here, we consider co-propagating pump and signal waves. For a counter-propagating pump and signal configuration, the sign of $g_p$ in Eq. (\ref{PumpPower}) should be reversed. Pump absorption coefficient is given by
\begin{equation} \label{PumpGain}
g_p = \Gamma_p \left[ \sigma_e \left( \omega_p \right) n_2- \sigma_a \left( \omega_p \right) n_1 \right].
\end{equation}
Power spectral densities of forward $\rho^f$ and backward $\rho^b$ propagating ASE are given \cite{Lindberg} by
\begin{equation} 
  \pm \partial_z \rho^{f(b)}(\omega) =  \bigl[(\beta_{\omega} + \mu_{\omega}) n_2 -\mu_{\omega} n_{\rm tot} \bigr] \rho^{f(b)}(\omega)+2 \beta_{\omega} n_2  \hbar \left( \omega_s+\omega \right)/\left( 2 \pi \right), \label{ro}
\end{equation}
where $\beta_{\omega}=\Gamma_s \sigma_e(\omega)$ and $\mu_{\omega}=\Gamma_s \sigma_a(\omega)$, and $+(-)$ should be chosen for the forward (backward) propagating ASE. Since Eq. (\ref{GNLSE}) is written in the co-moving frame, $\rho(\omega)$ and $\sigma \left( \omega \right)$ should formally be interpreted as $\rho(\omega+\omega_s)$ and $\sigma ( \omega +\omega_s )$, respectively. For brevity, the shift by $\omega_s$ is omitted throughout the paper. Higher-order nonlinear effects, namely self-steepening and stimulated Raman scattering (SRS), are neglected in Eq. (\ref{GNLSE}). Throughout the amplification process in the GMN regime, the pulse remains strongly positively chirped with a picosecond-scale temporal duration and a fractional spectral bandwidth $\delta \omega / \omega_s \lesssim 0.05$, rendering self-steepening negligible \cite{Lindberg}, \cite{Kozlov}. Furthermore, for the operating parameters considered in this work, the amplified pulse energies remain below the SRS threshold of the $6 \rm \mu m$ -core fiber \cite{Sidor}, so that spectral evolution is governed primarily by the interplay of self-phase modulation, group-velocity dispersion, and frequency-dependent gain.

\subsection{Unperturbed Solution}

For experimentally relevant parameters, the recovery time of the inversion 
\newline $t_c=1/\left[R^p_e + R^p_a+1/\tau_{ul}\right]$ \cite{Lindberg} is much greater than $T_{\rm rep}$ and therefore the pump will not be able to recover  the inversion between consecutive pulses. In this case, the time derivative of the inversion in Eq. (\ref{Rateq2}) can be neglected, yielding unperturbed inversion:
\begin{equation} \label{Rateq3}
    n^0_2 =\frac{n_{\rm tot} R_a}{ R_a+R_e+1/\tau_{ul}} ,
\end{equation}
The propagation dynamics of the unperturbed pulse, characterized by $\psi_0$, $P_{p0}$, and $\rho^{f(b)}_0$, were evaluated by numerically solving Eqs. (\ref{GNLSE})–(\ref{Gain}) and (\ref{total_rate})–(\ref{Rateq3}), with $\psi$, $P_p$, $\rho^{f(b)}$, and $n_2$ replaced by their corresponding unperturbed values. The numerical procedure is iterative: each iteration consists of a forward propagation step to evaluate the signal and forward ASE PSD, followed by a backward propagation step to evaluate the backward ASE PSD. The GNLSE (\ref{GNLSE}) was solved using the fourth-order Runge–Kutta interaction picture (RK4IP) method \cite{Lindberg, Hult}, which alternates between the Fourier domain for evaluating dispersion and the time domain for evaluating nonlinear effects.

\subsection{Pump and Seed Noise}
Coupled equations for positive $\delta \psi_+$ and negative $\delta \psi_-$ frequency components of the perturbation of the envelope of the electric field were obtained by substituting $\psi = \psi_0 + \delta \psi_+ \exp(i 2 \pi \nu T) + \delta \psi_- \exp(-i 2 \pi \nu T)$ into Eq. (\ref{GNLSE}) ( $T$ is slow pulse-to-pulse time), linearizing the result and keeping only resonant terms 
\begin{align}\label{perturbation}
\partial_z \delta \psi_+ &= -i \beta_2 \partial_{tt} \delta \psi_+/2 + i 2 \gamma |\psi_0|^2 \delta \psi_+ + i \gamma \psi_0^2 \delta \psi_-^* + \mathcal{F}^{-1} \{ g_0 \delta \tilde{\psi_+} + \delta g \tilde{\psi_0}\}/2, \\
\partial_z \delta \psi_- &= -i \beta_2 \partial_{tt} \delta \psi_-/2 + i 2 \gamma |\psi_0|^2 \delta \psi_- + i \gamma \psi_0^2 \delta \psi_+^* + \mathcal{F}^{-1} \{ g_0 \delta \tilde{\psi_-} + \delta g^* \tilde{\psi_0}\}/2.
\end{align}
should be solved with $\psi_0$ and $g_0$ found from Eqs. (\ref{GNLSE}-\ref{ro}). Perturbation of wave function and perturbation of gain $\delta g$ depend also on radio frequency $\nu$: $\delta \psi_{\pm}= \delta \psi_{\pm}(t, z, \nu)$, $\delta g = \delta g (\omega, z, \nu)$. Fourier transform of perturbation of wave function is given by $\delta \tilde \psi_{\pm}\left( \omega,z, \nu \right)=\mathcal{F} \{  \delta \psi_{\pm}(t,z, \nu) \}$. By linearizing Eqs. (\ref{Gain}--\ref{ro}) one finds equations for perturbation of gain:
\begin{equation}
    \delta g = \delta n_2 \Gamma_s \left[ \sigma_e(\omega)+\sigma_a(\omega)\right];
\end{equation}
perturbation of inversion: 
\begin{equation}
  \delta n_2 = \frac{n^0_1 \delta R_a - n^0_2 \delta R_e}{D} , 
\end{equation}
where 
\begin{equation}
    D= 2 \pi i \nu +1/\tau_{ul} + R^0_a + R^0_e;
\end{equation}
perturbations of emission (absorption) rates:
\begin{equation}
  \delta R_{e(a)} = \delta R^s_{e(a)} + \delta R^p_{e(a)} + \delta R^{f\ ase}_{e(a)} + \delta R^{b\ ase}_{e(a)}; 
\end{equation}
perturbations of signal emission (absorption) rates:
\begin{equation}
  \delta R^s _{e(a)} = \frac{\Gamma_s}{T_{\rm rep} \hbar A_d} \int \frac {\sigma_{e(a)} \left( \omega \right)\left[\tilde \psi_0^*\left( \omega \right) \delta \tilde \psi_+\left( \omega \right) + \tilde \psi_0\left( \omega \right) \delta \tilde \psi_-^*\left( \omega \right) \right]} {\omega_s+\omega} d\omega; \label{dRatese}
\end{equation}
perturbations of pump emission (absorption) rates:
\begin{equation}
    \delta R^p_{e(a)} = q_p R^p_{e(a)};
\end{equation}

relative perturbation of pump power ($q_p(\nu, z) = \delta P_p(\nu, z)/P_p^0(z) $) :
\begin{equation}
   \partial_z q_p = \delta n_2 \Gamma_p \left[ \sigma_e(\omega_p)+\sigma_a(\omega_p)\right] ;
\end{equation}
perturbation of forward (backward) ASE emission and absorption rates:
\begin{equation}
  \delta R^{f(b)} _e = \frac{1}{ \hbar A_d} \int \frac {\beta_{\omega} \delta \rho^{f(b)}} {\omega_s+\omega} d\omega . \label{dRoe}
\end{equation}
and 
\begin{equation}
  \delta R^{f(b)} _a = \frac{1}{ \hbar A_d} \int \frac {\mu_{\omega} \delta \rho^{f(b)}} {\omega_s+\omega} d\omega; \label{dRoa}
\end{equation}
perturbations of power spectral densities of forward (backward) ASE:
\begin{equation}
 \begin{split}
   \pm \partial_z \delta \rho^{f(b)}(\omega,\nu) &= \bigl[(\beta_{\omega} + \mu_{\omega}) n^0_2 -\mu_{\omega} n_{\rm tot} \bigr] \delta \rho^{f(b)}(\omega) \\ 
   &+\left[ 2 \beta_{\omega} \hbar \left( \omega_s+\omega \right)/\left( 2 \pi \right) + \left(\beta_{\omega} + \mu_{\omega}\right) \rho_0 \right] \delta n_2, \label{drof}   
\end{split}    
\end{equation}
where $+(-)$ should be chosen for the forward (backward) propagating ASE.

System of equations (\ref{perturbation}-\ref{drof}) was evaluated with two different initial conditions: \\
1) $q_p(\nu, z=0)=0$ and $\delta \psi_+(z=0)=\delta \psi_-(z=0)=\frac{1}{2} \psi_0(z=0)$, $\delta \rho^f(z=0)=0$, $\delta \rho^b(z=L)=0$ for signal-to-signal and signal-to-ASE noise transfer functions (NTF), and \\
2) $q_p(\nu, z=0)=1$ and $\delta \psi_{\pm}(z=0)=0$, $\delta \rho^f(z=0)=0$, $\delta \rho^b(z=L)=0$ for pump-to-signal and pump-to-ASE NTF.

The system of eqs. (\ref{perturbation}--\ref{drof}) was solved numerically using the fourth-order Runge–Kutta interaction picture (RK4IP) method \cite{Lindberg, Hult}. The numerical procedure is iterative: each iteration consists of a forward propagation step to evaluate the perturbations of the signal and forward ASE PSD, followed by a backward propagation step to evaluate the perturbation of backward ASE PSD . 

Signal-to-signal $\xi^s_{s}(\nu)$ and pump-to-signal $\xi^s_{p}(\nu)$ noise transfer functions (NTF) can be found by $\delta E / E^s_0$, where signal energy 
\begin{equation}
 E^s_0 = \int |\psi_0|^2 dt = \int |\tilde \psi_0|^2 d \omega, \label{SE}    
\end{equation}

and perturbation of signal energy 
\begin{equation}
 \delta E = \int (\psi_0^* \delta \psi_+ + \psi_0 \delta \psi_-^*) dt = \int (\tilde \psi_0^* \delta \tilde \psi_+ + \tilde \psi_0 \delta \tilde \psi_-^*)d \omega  , \label{PSE} 
\end{equation}
 are to be found by solution of system (\ref{perturbation}-\ref{drof}) with corresponding initial conditions.

Signal-to-ASE $\xi^{ASE}_{s}$ and pump-to-ASE $\xi^{ASE}_{p}$ NTF can be found by 
\begin{equation}
   \xi^{ASE}_{s(p)} (\nu) = \int  \delta \rho^{f} (\omega, \nu) d\omega / P^s_0, \label{dro_f} 
\end{equation}

where signal power $P^s_0=E^s_0/T_{\rm rep}$ and $\delta \rho^{f} (\omega, \nu)$ is perturbation of the forward ASE PSD at $z=L$ with corresponding initial conditions.

The spectral density of relative intensity noise RIN (RSD) of the amplifier output signal noise due to pump and seed noises (technical noise) is given by:
\begin{equation}
    S_{\rm tech}(\nu)=\sum _{k=p,s} \bigl| \xi_k^s(\nu)+\xi_k^{ASE}(\nu)\bigr|^2 S_{k}(\nu), \label{samp}
\end{equation}
where $S_{p}(\nu)$ and $S_{s}(\nu)$ are input pump and signal (seed) RSDs, respectively, and we used the fact that fluctuations of pump and seed are uncorrelated. The corresponding RIN (in percent) is given by
\begin{equation}
    \mathrm{RIN_{tech}}=100 \frac{P^s_0}{P^s_0 + P_0^{ASE}} \sqrt{\int_{\nu_{\rm min}}^{\nu_{\rm max}} S_{\rm tech}(\nu)   d\nu}\label{RINtech},
\end{equation}
where 
\begin{equation}
    P_0^{ASE} = \int  \rho_0^f (\omega) d \omega, \label{P0ASE}
\end{equation}
and $\nu_{\rm max}=min(f_{\rm rep}/2, f_{\rm max \ det})$, $\nu_{\rm min}= f_{\rm min \ det}$ and $f_{\rm max \ det}$ and $f_{\rm min \ det}$ are upper and lower limits of detector bandwidth, respectively. The NTFs in the above expressions are normalized to the mean signal power $P_{s}^0$, in which form they have a direct physical meaning ($\xi_s^s \rightarrow 1$ corresponds to unity transfer of seed fluctuations, and $\xi_p^s(0)$ is the logarithmic slope of the signal power with respect to the pump power) and coincide with the analytical expressions of Eqs. (\ref{xis})–(\ref{xip}). The RIN, however, is defined relative to the total detected output power $P_{s}^0 + P_{ASE}^0$, which is what a photodetector without spectral filtering measures. Since both $P_s$ and $P_{ASE}^0$ are mean values, the conversion between the two normalizations is the frequency-independent factor $P_{s}^0/(P_{s}^0 + P_{ASE}^0)$ in Eq. (\ref{RINtech}). The same frequency-independent factor applies to the signal-ASE and ASE-ASE beat noise and to the shot noise, so all RIN components in the paper refer to the same mean power. Importantly, this approach enables the evaluation of the output signal RSD by solving Eqs. (\ref{GNLSE})--(\ref{drof}) only once, thereby avoiding computationally intensive statistical simulations involving random noise terms.

\subsection{ASE Noise}

It is well known \cite{olson, Beker, hui, desert} that amplified spontaneous emission gives rise to slow (RF) power fluctuations due to signal–ASE and ASE–ASE beat noise. The expressions for signal–ASE and ASE–ASE beat noise derived in the above references are based on the monochromatic approximation of the signal, with its frequency assumed to coincide with the central frequency of the ASE spectrum, and the approximation that the ASE spectrum is rectangular. In the present work, we generalize these expressions by taking into account the exact spectra of both the signal and ASE. Accounting for the exact signal spectrum is particularly important for amplifiers operating in the gain-managed nonlinearity regime, where the finite gain bandwidth and spectral redshift \cite{Sidor, Kozlov} become significant. Henceforth, we omit the superscript $^f$ in $\rho^f$ for brevity. The expressions for RSDs of signal-ASE beat noise
\begin{equation}
   S_{\rm S-ASE}(\nu) = \frac{2 \pi }{T_{\rm rep}\left(P_s^0\right)^2} \int |\tilde{\psi}\left( \omega \right)|^2 \left[ \rho_0 (\omega-2 \pi \nu)+\rho_0 (\omega+2 \pi \nu) \right] d\omega \label{Signal-ASE}
 \end{equation}
and ASE-ASE beat noise 
\begin{equation}
   S_{\rm ASE-ASE}(\nu) = \frac{2 \pi }{\left(P_s^0\right)^2} \int \rho_0 (\omega)\rho_0 (2 \pi \nu+\omega) d\omega \label{ASE-ASE}
 \end{equation}
are derived in Appendix A. Eq. (\ref{ASE-ASE}) is consistent with corresponding equations given in \cite{olson, Beker, hui, desert}. The RIN contributions of the signal–ASE and ASE–ASE beat noises are obtained from Eq. (\ref{RINtech}) by substituting the corresponding RSD.

\subsection{Shot Noise}
The RSD of the shot noise was evaluated by using well known \cite{Beker, desert} expression 
\begin{equation}
    S_{\rm shot}=\frac{2 \hbar \omega_s \left( P_{s}^0 + P_{ASE}^0\right)}{{P_{s}^0}^2}, \label{shot}
\end{equation}
Due to the red shift of the pulse in the GMN regime the photon energy is $3 - 5 \%$ lower than at $1030 {\rm nm}$, so the shot-noise RIN, which scales as $\sqrt{\hbar \omega}$, is overestimated by negligible $2 - 3 \%$. The RIN contribution of the shot noise is obtained from Eq. (\ref{RINtech}) by substituting the corresponding RSD.

\section{Analytical Solution of Monochromatic Approximation}

Our analytical model is based on the rate-equation perturbation method originally developed for EDFAs in \cite{Novak} and subsequently applied to Yb-doped fiber amplifiers in \cite{Zhao2017, Gierschke2019, Han2021}, which we adapt here for a two-level gain medium and extend to include the signal- and pump-to-ASE noise transfer functions. In the monochromatic approximation we neglect finite bandwidth of the signal so the signal gain coefficient is given by
\begin{equation} \label{PumpGain}
g_s = \Gamma_s \left[ \sigma_e \left( 0 \right) n_2- \sigma_a \left( 0 \right) n_1 \right].
\end{equation}
Signal power then can be evaluated by
\begin{equation} \label{SignalPowerA}
d_z {P_s}= g_s P_s,
\end{equation}
Signal emission (absorption) rates in monochromatic approximation are given by
\begin{equation}
    R^s_{e(a)}=\frac{\Gamma_s\sigma_{e(a)} \left( 0 \right) P_s}{\hbar A_d \omega_s}.
\end{equation}
For our analytical model we assume that $P_{\rm ASE}\ll P_s$. In this case, the depletion of the excited state by forward and backward ASE can be neglected and Eq. (\ref{total_rate}) can be approximated by 
 \begin{equation} \label{total_rateM}
R_{e(a)}=R^s_{e(a)}+R^p_{e(a)}.
\end{equation}   
Eq. (\ref{Rateq2}) then becomes:
\begin{equation} \label{RateqA}
    \frac {\partial n_2  }{\partial t}= -\sum_{k=p,s}\frac{g_k P_k}{\hbar \omega_k A_d}-n_2/\tau_{ul}=-\sum_{k=p,s}\frac{d_z P_k}{\hbar \omega_k A_d}-n_2/\tau_{ul}.
\end{equation}
By integrating Eq. (\ref{RateqA}) over the length of fiber, one finds that: 
\begin{equation}
   \frac{\partial N_2}{\partial t} = \sum _{k=p,s}\frac{P_k(0)-P_k(z)}{\hbar \omega_k}  - \frac{N_2}{\tau_{ul}}\label{N2} ,
\end{equation}
where $N_2=A_d \int_0^z n_2(\zeta) d\zeta$ is the total number of ions in the excited state. By integrating Eqs. (\ref{PumpPower}) and (\ref{SignalPowerA}), one finds  that
\begin{equation}
    P_k\bigl(z \bigr) =P_k \bigl( 0 \bigr) \exp\left( B_k N_2 - C_k\right)\label{Rtesn} ,
\end{equation}
where dimensionless quantities $B_k=\Gamma_k \left[ \sigma_{e} \left( \omega_k \right) + \sigma_{a}\left( \omega_k \right) \right] / A_d$ and $C_k=\Gamma_k \sigma_{a}\left( \omega_k \right) n_{\rm tot} z  $. 

\subsection{Unperturbed Solution}
Steady-state solution $N_2^0=A_d \int_0^z n_2^0(\zeta) d\zeta$  can be found by numerical solution of the transcendental equation
\begin{equation}
   \frac{N_2^0}{\tau_{ul}} = \sum _{k=p,s}\frac{P_k^0(0)}{\hbar \omega_k} \bigl[ 1-\exp\left( B_k N_2^0 - C_k\right) \bigr]  \label{N20} ,
\end{equation}
obtained from Eq. (\ref{N2}) by setting ${\partial N_2}/{\partial t}$ to zero and substituting the steady-state powers 
\begin{equation}
    P_k^0\bigl(z \bigr) =P_k^0 \bigl( 0 \bigr) \exp\left( B_k N_2^0 - C_k\right)\label{Rtes0}.
\end{equation}

\subsection{Pump and Seed Noise}
On the next step we introduce small ($q_i\ll1$) frequency-dependent modulations of the amplitudes:
\begin{align}
     N_2(z,t) & = N_2^0(z)\left[1+\left(q_n(\nu, z) e^{i 2 \pi \nu t}+ c.c. \right) \right],\label{fouriern} \\ 
     P_{p(s)}(z,t) & = P_{p(s)}^0(z)\left[1+\left(q_{p(s)} (\nu, z) e^{i 2 \pi \nu t} + c.c.\right)\right]\label{fourierP}.
\end{align}
By substituting Eq. (\ref{fouriern}) and (\ref{fourierP}) into Eqs. (\ref{N2}) and (\ref{Rtesn}) and linearizing them with respect to small power modulations, one finds that
\begin{align}
\begin{split}
     i 2 \pi \nu N_2^0 q_n(\nu, z) & = \sum _{k=p,s} q_k (\nu, 0) 
    \frac{P_k ^0 (0)-P_k ^0 (z)} {\hbar \omega_k} \\
    & - \frac{N_2^0 q_n(\nu, z) B_k P_k ^0 (z)}{\hbar \omega_k} -\frac{N_2^0 q_n(\nu, z)}{\tau_{ul}} \label{n2q} 
    \end{split} \\
    q_s(\nu, z) & = q_s(\nu,0) + N_2^0 q_n(\nu, z) B_s. \label{qS}
\end{align}


By expressing $N_2^0 q_n(\nu, z)$ in terms of $q_p(\nu,0)$ and $q_s(\nu,0)$, using Eq. (\ref{n2q}) and substituting the resulting expressions into Eq. (\ref{qS}), one finds that the signal- and pump-to-signal NTFs are
\begin{align}
    & \xi^s_s(\nu) = 1 + \frac {B_s X_s}{D}, \label{xis} \\ 
    & \xi^s_p(\nu) = \frac{B_s X_p}{D}\label{xip},
\end{align}
where $X_k(z)=\bigl[P_k^0 (0) - P_k^0 (z) \bigr]/(\hbar \omega_k)$ and $D(z)=\sum _{k=p,s} B_k P_k^0(z)/(\hbar \omega_k) + i 2 \pi \nu + 1/\tau_{ul}$. 

If the depletion of the excited state by ASE can be neglected, then Eq. (\ref{ro}) decouples from Eqs. (\ref{Rateq2})--(\ref{PumpPower}) \cite{Jopson}.
%
In this case Eq. (\ref{ro}) has the following analytical solution for PSD of forward propagating ASE
\begin{equation}
\begin{split}
  \rho^f(\omega) & = 2 \beta_{\omega} \hbar \left( \omega_s + \omega \right) / (2 \pi) \\
  & \times \int_0^z d \zeta n_2 (\zeta)  \exp{\bigl[\bigl(\beta_{\omega} + \mu_{\omega}\bigr)\frac{N_2(z)-N_2(\zeta)}{A_d} - \mu_{\omega} n_{\rm tot}\bigl(z-\zeta\bigr) \bigr]} , \label{ros}
  \end{split}
\end{equation}
and steady-state solution is given by
\begin{equation}
\begin{split}  
  \rho_0^f(\omega) & = 2 \beta_{\omega} \hbar \left( \omega_s + \omega \right) / (2 \pi) \\
  & \times \int_0^z d \zeta n_2^0 (\zeta)  \exp{\bigl[\bigl(\beta_{\omega} + \mu_{\omega}\bigr)\frac{N_2^0(z)-N_2^0(\zeta)}{A_d} - \mu_{\omega} n_{\rm tot}\bigl(z-\zeta\bigr) \bigr]} . \label{ros0}
\end{split}
\end{equation}
Perturbations of PSD of forward propagating ASE were obtained by substituting Eq. (\ref{n2q}) into Eq. (\ref{ros}) and linearizing the resulting equation with respect to small power modulations $q_k$:
\begin{equation}
\begin{split}
  \delta \rho^f (\omega, \nu) & = \frac{2 \beta_{\omega} \hbar (\omega_s+\omega)}{(2 \pi)} \\ & \times \biggl\{\int_0^z d \zeta 
  \biggl[ n_2^0 (\zeta) \frac{\beta_{\omega}+\mu_{\omega}}{ A_d} \biggl(\frac{X_k(z)}{D(z)} - \frac{X_k(\zeta)}{D(\zeta)} \biggr)   + \frac{1}{A_d}\frac{d}{d \zeta} \frac{X_k(\zeta)}{D(\zeta)}\biggr] \\
 & \times \exp{\bigl[\bigl(\beta_{\omega} + \mu_{\omega}\bigr)\frac{N_2^0(z)-N_2^0(\zeta)}{A_d} - \mu_{\omega} n_{\rm tot}\bigl(z-\zeta\bigr) \bigr]} \biggr\} , \label{K_ASE}
\end{split}
\end{equation}
where $k=p$ for pump-to-ASE NTF and $k=s$ for signal-to-ASE NTF. To verify Eq. (\ref{K_ASE}), we derived it independently by using an alternative approach, presented in Appendix B. Eqs. (\ref{dro_f}-\ref{RINtech}) remain unchanged.

\subsection{ASE Noise}
In monochromatic approximation Eq. (\ref{Signal-ASE}) becomes
\begin{equation}
  S_{\rm S-ASE}(\nu) = \frac{ 2 \pi  \left[\rho_0 \left( 2 \pi \nu \right)+\rho_0 \left( -2 \pi \nu \right) \right]}{P_s^0  }.  \label{S_ASE}
\end{equation}
Equation (\ref{S_ASE}) is consistent with corresponding equations given in \cite{olson, Beker, hui, desert}. Expression for the RSD of ASE-ASE beat noise is given by Eq. (\ref{ASE-ASE}). The RIN contributions of the signal–ASE and ASE–ASE beat noises are obtained from Eq. (\ref{RINtech}) by substituting the corresponding RSD.

\section{Results and Discussion}
In the following we present results obtained by these numerical and analytical methods for parameters listed in the Table \ref{table1}. 
\begin{table}[h!]
\centering
\caption{Fiber Laser Parameters}
\begin{tabular}{ |c|c|c|c|c|c|} 
 \hline
 $\beta_2 \bigl[\rm ps^2/m\bigr]$ & $\gamma  \bigl[1/(\rm W \times m)\bigr]$ & $\tau_{ul} \bigl[\rm s\bigr]$ & {$n_{\rm tot} \bigl[\rm m^{-3}\bigr]$} & $d_{\rm core} \bigl[\rm m\bigr]$ & $d_{\rm clad} \bigl[\rm m\bigr]$ \\
 \hline
 $0.024$ & $0.0038$ & $10^{-3}$ & {$0.814 \times 10^{26}$} & $6 \times 10^{-6}$ & $1.25 \times 10^{-4}$\\
  \hline\end{tabular}
\label{table1}
\end{table}
 These parameters correspond to a typical fiber amplifier based on a commercially available double-clad Yb-doped fiber (YB1200-6/125DC-PM). The signal carrier wavelength is $\lambda_s=2 \pi c/\omega_s = 1030 \ \rm nm$ and the pump wavelength is $\lambda_p=2 \pi c/\omega_p = 976 \ \rm nm$. Since a cladding-pumped amplifier is considered, the pump overlap factor with the doped core region is given by $\Gamma_p=A_d/A_{\rm clad}=0.0023$, where $A_{\rm clad}=\pi (d_{\rm clad}/2)^2$. The signal overlap factor with the doped core region is $\Gamma_s=A_d/A_{\rm eff}=0.7266$, where the effective mode area is $A_{\rm eff}=\pi r_{\rm eff}^2$. The effective mode radius is calculated using the empirical expression \cite{Marcuse} $r_{\rm eff} = (0.65+1.619 \ V^{-3/2}+2.879 \ V^{-6}) \times d_{\rm core}/2$, where $V={\rm NA} \times  2 \pi d_{\rm core}/(2\lambda_s)$ and numerical aperture ${\rm NA}=0.12$.

\subsection{Linear Low-ASE Regime}
As a first step, the model is validated by comparing the results of the exact numerical simulations with those obtained using the monochromatic analytical approximation in a regime where both excited-state depletion by ASE and NSB are negligible. To satisfy these conditions, a relatively long seed pulse ($\mathrm{FWHM}=16  \ \mathrm{ps}$) is used to suppress nonlinear effects, while a relatively high input pulse energy ($E_s(z=0)=2 \  \mathrm{nJ}$) is employed to minimize the influence of excited-state depletion by ASE. For this comparison we used pump power $P_p=0.2  \ \rm W$, length of fiber $L_{\rm fiber}=3.5 \ \rm m$ and repetition rate $f_{\rm rep}=1/T_{\rm rep}=4 \ \rm M\rm Hz$.

\begin{figure} [b]
    \begin{center}
         \includegraphics[width=130mm]{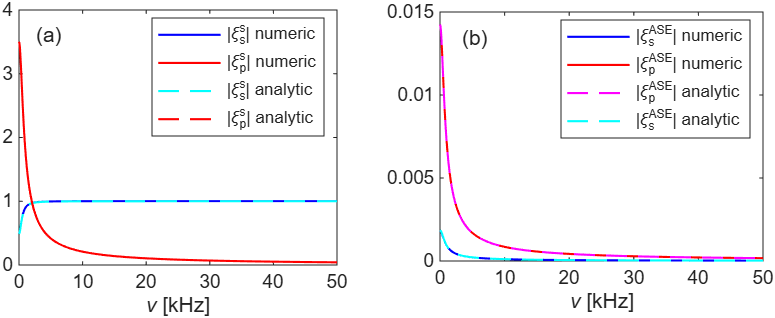}
    \caption{The signal-to-signal and pump-to-signal NTFs (a) and signal-to-ASE and pump-to-ASE NTFs (b) versus RF. The numerical pump-to-signal and pump-to-ASE NTFs are shown by red solid curves, and the numerical signal-to-signal and signal-to-ASE NTFs are shown by blue solid curves. The analytical pump-to-signal and pump-to-ASE NTFs are represented by magenta dashed curves, and the analytical signal-to-signal and signal-to-ASE NTFs are represented by cyan dashed curves.}
     \label{Fig1}
    \end{center}
\end{figure}

The NTFs as a function of RF are presented in Fig. \ref{Fig1}. Figure \ref{Fig1}(a) shows the signal-to-signal and pump-to-signal NTFs, whereas Fig. \ref{Fig1}(b) presents the signal-to-ASE and pump-to-ASE NTFs. The numerical pump-to-signal and pump-to-ASE NTFs are shown by red solid curves, whereas the numerical signal-to-signal and signal-to-ASE NTFs are shown by blue solid curves. The corresponding analytical results for the pump-to-signal and pump-to-ASE NTFs are represented by magenta dashed curves, while the analytical signal-to-signal and signal-to-ASE NTFs are represented by cyan dashed curves. These figures demonstrate excellent agreement between the analytical and numerical results: the two sets of results are visually indistinguishable. Importantly, the signal-to-signal NTF increases with frequency, whereas the pump-to-signal NTF decreases with frequency. This behavior can be explained as follows. Both pump and seed noise affect the output signal noise through fluctuations in the excited-state population, as represented by the terms ($X_k/D$) in Eqs. (\ref{xis}) and (\ref{xip}). Due to the finite response time (inertia) of the system, these contributions decrease with increasing frequency. In contrast, seed noise is also transferred directly to the output signal, as represented by the unity term in Eq. (\ref{xis}). These two mechanisms act in opposition: an increase in the seed field leads to a reduction in the excited-state population, resulting in partial cancellation of the corresponding fluctuations. At low frequencies, where the system response is strongest, this cancellation is most effective, leading to a relatively small signal-to-signal NTF. As the frequency increases, the system inertia suppresses fluctuations of the excited-state population, thereby reducing the compensating effect. Consequently, the direct transfer of seed-field fluctuations becomes increasingly dominant, resulting in a larger signal-to-signal NTF.

The dependence of the signal–ASE and ASE–ASE beat-noise RSDs on the RF ($\nu$) was neglected, as the ASE PSD remains nearly constant over the considered radio-frequency range. The numerically calculated RSD of the signal–ASE beat noise is $S_{\rm S-ASE}=3.63 \times 10^{-16} 1/\rm Hz$, whereas the corresponding analytical value is $S_{\rm S-ASE}=3.63 \times 10^{-16} \ 1/\rm Hz$. The numerically calculated RSD of the ASE–ASE beat noise is $S_{\rm ASE-ASE}=4.907 \times 10^{-18} \ 1/\rm Hz$, while the corresponding analytical value is $S_{\rm ASE-ASE}=4.89 \times 10^{-18} \ 1/\rm Hz$. The numerically calculated RSD of the shot noise is $S_{\rm shot}=1.353 \times 10^{-17} \ 1/\rm Hz$, and the corresponding analytical value is $S_{\rm shot}=1.341 \times 10^{-17} \ 1/\rm Hz$. 
We further consider two different nonlinear regimes. 

\subsection{High-power, High-repetition-rate Nonlinear Regime}
\begin{figure} [!b]
    \begin{center}
    \hspace*{-1.5 cm}
         \includegraphics[width=140mm]{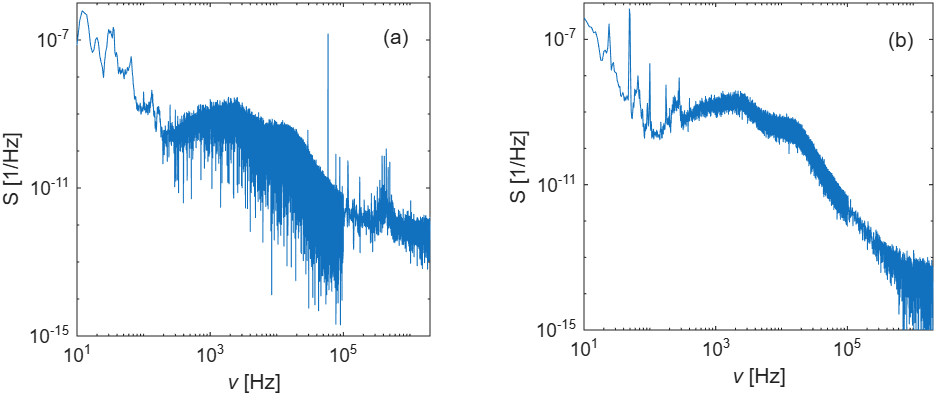}
    \caption{RSDs of the pump (a) and seed (b) for the high-power, high-repetition-rate regime \cite{Nitzan}.}
     \label{Fig2}
    \end{center}
\end{figure}

In the first high-power, high-repetition-rate regime, we used a seed pulse with energy $E_s(z=0)=0.2 \ \rm nJ$ and $\rm FWHM=0.8 \ ps$. The pump power was varied between $2.5$ and $16 \ \rm W$ and the repetition rate between $30$ and $80 \ \rm MHz$. These parameters place the amplifier in the GMN regime, which is characterized by significant NSB and red shift of the signal spectrum, as demonstrated below. For the RSDs of the pump and seed, presented respectively in Fig. \ref{Fig2} (a) and \ref{Fig2} (b), we used the experimental data given in \cite{Nitzan}.

\begin{figure} 
    \begin{center}
    \hspace*{-1.5 cm}
         \includegraphics[width=140mm]{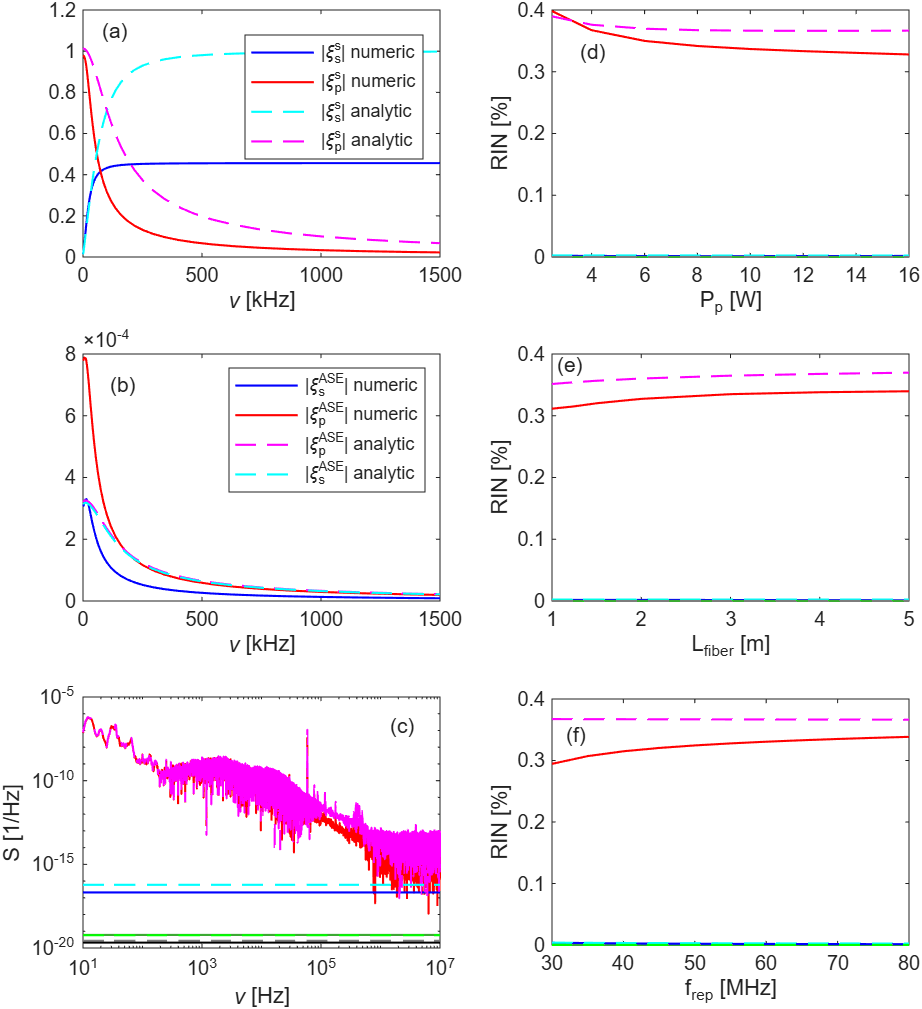}
    \caption{Results for the high-power, high-repetition-rate regime: (a) signal-to-signal and pump-to-signal NTFs, (b) signal-to-ASE and pump-to-ASE NTFs, (c) RSDs of the different noise sources, all plotted as functions of RF for $P_p=10 \ \rm W$, $f_{\rm rep}=75 \ \rm MHz$ and $L_{\rm fiber}=3.5 \rm m$. In (a-b), the numerical pump-to-signal and pump-to-ASE NTFs are shown as solid red curves, whereas the numerical signal-to-signal and signal-to-ASE NTFs are shown as solid blue curves. The analytical pump-to-signal and pump-to-ASE NTFs are shown by dashed magenta curves, and the analytical signal-to-signal and signal-to-ASE NTFs are shown by dashed cyan curves. In (d-f), the RINs of different noise sources are plotted as functions of pump power for $L_{\rm fiber}=3.5 \rm m$ and $f_{\rm rep}=75 \ \rm MHz$ (d), as functions of fiber length for $P_p=10 \rm W$ and $f_{\rm rep}=75 \ \rm MHz$ (e), and as functions of repetition rate for $P_p=10 \rm W$ and $L_{\rm fiber}=3.5 \rm m$ (f). In (c-f), the solid red curves represent numerically calculated results for the technical noise, and the dashed magenta curves represent the corresponding analytical results. The signal–ASE beat noise is shown by the solid blue (numerical) and dashed cyan (analytical) curves. The ASE–ASE beat noise is represented by the solid black (numerical) and dashed gray (analytical) curves. The shot noise is shown by solid dark green (numerical) and dashed light green (analytical) curves.}
     \label{Fig3}
    \end{center}
\end{figure}

\begin{figure} [t]
    \begin{center}
    \hspace*{-1.5 cm}
         \includegraphics[width=140mm]{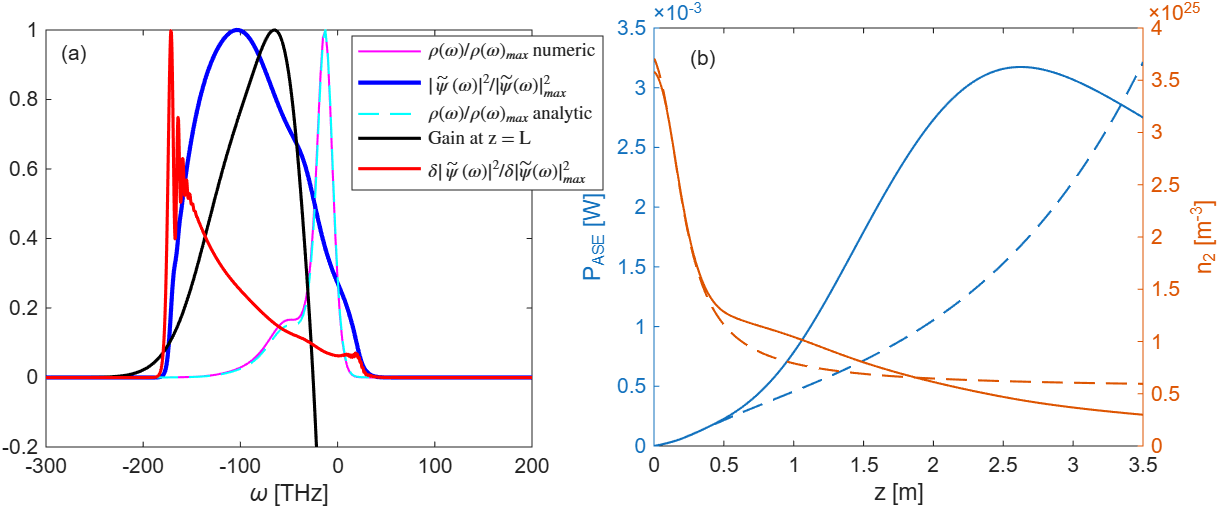}
    \caption{(a) Signal spectrum (solid blue curve) and perturbation of signal (solid red curve) spectrum for $\nu=4 \rm MHz$; Numerical (solid magenta curve) and analytical (dashed cyan curve) ASE PSD. All values are at the fiber output and normalized to their maximum values. (b) Numerical (solid blue curve) and analytical (dashed blue curve) ASE powers, and numerical (solid red curve) and analytical (dashed red curve)  inversion versus propagation length.}
     \label{Fig4}
    \end{center}
\end{figure}

Figure \ref{Fig3}(a) shows the signal-to-signal and pump-to-signal NTFs, Fig. \ref{Fig3}(b) presents the signal-to-ASE and pump-to-ASE NTFs, and Fig. \ref{Fig3}(c) depicts the RSDs of the different noise sources, all plotted as functions of RF for $P_p=10 \ \rm W$, $f_{\rm rep}=75 \ \rm MHz$ and $L_{\rm fiber}=3.5 \ \rm m$. In Fig. \ref{Fig3} (a) and \ref{Fig3} (b) the numerical pump-to-signal and pump-to-ASE NTFs are shown as solid red curves, whereas the numerical signal-to-signal and signal-to-ASE NTFs are shown as solid blue curves. The corresponding analytical results are represented by dashed magenta and dashed cyan curves, respectively. The most striking discrepancy between the analytical and numerical results is the asymptotic behavior of the signal-to-signal NTF at high modulation frequencies. The analytical NTF approaches unity because the inversion cannot respond to rapid seed fluctuations due to its finite response time. Consequently, gain dynamics cannot suppress these fluctuations, which are therefore transferred unchanged to the output signal. 

In contrast, the asymptotic value of the numerically calculated signal-to-signal NTF lies well below unity owing to the Kerr nonlinearity. The origin of this discrepancy is illustrated in Fig. \ref{Fig4}(a), which shows the signal spectrum (blue curve), the gain (black curve), and the perturbation of the signal spectrum $\delta |\tilde \psi(\omega)|^2=\tilde \psi_0^*(\omega) \delta \tilde \psi_+(\omega) + \tilde \psi_0 (\omega) \delta \tilde \psi_-^*(\omega)$ for $\nu=4 \rm MHz$ (red curve) at the fiber output, each normalized to its maximum value. The asymmetry of the gain, with absorption on the short-wavelength side and amplification on the long-wavelength side of the signal spectrum, converts the symmetric NSB into a red shift that grows with pulse energy. The nonlinear Kerr terms in the perturbation equations [Eqs. (\ref{perturbation})] make the NSB of the perturbation stronger than that of the signal itself: a slightly higher pulse energy produces stronger self-phase modulation and extends the spectrum further into its wings, so the perturbation is concentrated at the outer spectral edges, outside the region that contains most of the pulse energy. The spectral width of the perturbation therefore exceeds that of the signal, and the difference increases with the B-integral. Combined with the gain asymmetry, this shifts the perturbation spectrum further toward long wavelengths than the signal spectrum, as seen in Fig. \ref{Fig4}(a). Since the signal spectrum lies on the long-wavelength side of the gain peak, where the gain decreases with wavelength, the perturbation experiences weaker amplification than the signal. This effect accounts for the reduced high-frequency asymptotic value of the numerically calculated signal-to-signal NTF, which we refer to below as the Kerr factor.

The numerical pump-to-signal NTF agrees well with the analytical result at low modulation frequencies but deviates significantly from it at higher frequencies. This behavior can be explained as follows. The numerator of Eq. (\ref{xip}) describes the sensitivity of the signal gain to changes in the inversion, whereas the denominator represents the saturation rate. Both terms contain the signal contribution term ($B_s$). For a signal with finite spectral bandwidth, the effective value of ($B_s$) should be calculated using energy-weighted cross sections averaged over the actual signal spectrum. NSB reduces this effective value. At low modulation frequencies, the signal contribution dominates the denominator, so the reduction in ($B_s$) affects both the numerator and denominator and largely cancels in their ratio. Consequently, the numerical pump-to-signal NTF remains close to the analytical prediction. At high modulation frequencies, however, the frequency-dependent term dominates the denominator, making the NTF approximately proportional to ($B_s$). The reduction in the effective ($B_s$) therefore leads to a significantly lower numerical pump-to-signal NTF than the analytical prediction.

ASE power perturbations are driven by inversion perturbations, which are inversely proportional to the saturation rate given by the denominator of Eq. (\ref{xip}). At low modulation frequencies, the saturation rate is dominated by the signal contribution ($B_s$), whose effective value is reduced by NSB of the signal, as discussed above. This reduction in the saturation rate leads to larger inversion perturbations and, consequently, higher numerical pump-to-ASE NTF values compared with the analytical predictions. At higher modulation frequencies, the saturation rate is dominated by the frequency-dependent term, and the numerical NTF approaches the analytical result. The signal-to-ASE NTF exhibits different behavior. At low modulation frequencies, the increase in inversion perturbations resulting from the reduced saturation rate is compensated by the reduction in signal fluctuations due to Kerr nonlinearities, as discussed above. Consequently, the numerical NTF remains close to the analytical prediction. At higher modulation frequencies, the saturation rate is dominated by the frequency-dependent terms, whereas Kerr nonlinearities continue to suppress signal fluctuations. As a result, the numerical signal-to-ASE NTF is lower than the analytical prediction.

Figure \ref{Fig3}(c) presents the numerically calculated RSDs of the technical noise, evaluated using Eq. (\ref{samp}), shown by the solid red curve. The corresponding analytical prediction is represented by the dashed magenta curve.  The frequency-independent RSD of the signal–ASE beat noise is shown by the solid blue (numerical) and dashed cyan (analytical) lines. The RSD of the ASE–ASE beat noise is shown by the solid black (numerical) and dashed gray (analytical) lines, and that of the shot noise by the solid dark green (numerical) and dashed light green (analytical) lines.

The numerically calculated RSD of the ASE–ASE beat noise $\approx 2.1 \times 10^{-20} \ {\rm Hz}^{-1}$ is approximately 23 \% lower than the analytical value $\approx 2.7 \times 10^{-20} \ {\rm Hz}^{-1}$. This difference originates from the lower ASE power in the numerical model, as shown in Fig. \ref{Fig4}(b), which results from the lower inversion in the output section of the fiber, also shown in Fig. \ref{Fig4}(b). The nonlinearly broadened signal spectrum extends to wavelengths where the absorption cross section is small, so that the signal continues to extract gain at low inversion and depletes the inversion near the fiber end more strongly than a monochromatic signal. Depletion of the inversion by forward and backward ASE, which is neglected in the analytical model, contributes only marginally at this pump power, since the ASE power is three orders of magnitude below the signal power. The analytically calculated RSD of the signal-ASE beat noise $\approx 5.98 \times 10^{-17} \ {\rm Hz}^{-1}$ is higher than the corresponding numerical value $\approx 2.086 \times 10^{-17} \ {\rm Hz}^{-1}$ by approximately a factor of three, which is due to the significant red shift and NSB of the signal in GMN regime, which results in a smaller overlap of the ASE and signal spectra. This is demonstrated in Fig. \ref{Fig4}(a), which shows the numerically calculated ASE PSD (solid magenta curve), the signal spectrum (solid blue curve), and the analytically calculated ASE PSD (dashed cyan curve), all normalized to their maximum values. The RSD of the ASE–ASE beat noise is significantly lower than that of the signal–ASE beat noise because the ASE power is much lower than the signal power. The numerically calculated shot-noise RSD ($\approx 6.05 \times 10^{-20}\ {\rm Hz}^{-1}$) is approximately 5 \% higher than the analytical value ($\approx 5.77 \times 10^{-20}\ {\rm Hz}^{-1}$), because the numerical signal power is lower owing to NSB of the signal in combination with the frequency-dependent gain.  

Figures \ref{Fig3}(d-f) show the integrated relative intensity noises (RINs) associated with the different noise sources. These RINs were evaluated using Eq. (\ref{RINtech}) with corresponding RSDs, and $\nu_{\rm min}=10 {\rm Hz}$, $\nu_{\rm max}=10 {\rm MHz}$. The curve colors and styles follow the same convention as in Fig. \ref{Fig3}(c). These figures demonstrate that the technical noise is the dominant contribution to the output signal noise. The signal-ASE and ASE-ASE beat-noise and shot noises contributions are negligible in this regime because the ASE power remains much lower than the signal power under these high-power, high-repetition-rate operating conditions. 

Figure \ref{Fig3}(d) depicts RINs versus pump power for $f_{\rm rep}=75 \ \rm MHz$, $L_{\rm fiber}=3.5 \ \rm m$. The analytically calculated technical RIN is nearly independent of pump power, decreasing slightly below 5 W and increasing marginally above 10 W. This behavior reflects the competition between the pump and seed noise channels, which are dominated by the pump-to-signal and seed-to-signal NTFs; the pump-to-ASE and seed-to-ASE contributions are comparatively small. Increasing the pump power shifts both the rise of the seed-to-signal NTF and the roll-off of the pump-to-signal NTF toward higher modulation frequencies, so that the pump-noise contribution grows while the seed-noise contribution shrinks. Above 10 W these opposite trends nearly cancel, and the RIN becomes essentially constant. Below 10 W the decrease of the seed-noise contribution outweighs the increase of the pump-noise contribution, for two reasons. First, the seed RSD rises more steeply toward low modulation frequencies than the pump RSD, so that the seed channel gains weight once the NTF corner frequency drops below 10 kHz at low pump power. Second, a lower pump power reduces the signal power and hence both the numerator of the fraction in Eq. (\ref{xis}), which quantifies the sensitivity of the signal gain to inversion fluctuations, and its denominator, which represents the total saturation rate. Because the denominator also contains the fixed spontaneous decay rate $1/\tau_{ul}$ and the residual pump power $P_p(L)$, which decreases more slowly, the fraction decreases and the zero-frequency seed-to-signal NTF increases. 

In the numerical model, NSB additionally reduces the effective signal contribution term $B_s$, as discussed above. This effect lowers the numerical RIN below the analytical prediction at high pump powers, where the pump-noise contribution dominates. At lower pump powers, the nonlinear suppression of the pump-channel contribution becomes weaker, while the seed-channel contribution becomes increasingly significant.  NSB shifts the rise of the seed-to-signal NTF toward lower frequencies owing to the reduced effective signal contribution coefficient $B_s$. Below 3 W this  raises the numerically calculated RIN above the analytical value. With increasing pump power, this effect is outweighed by three concurrent factors: the reduction of the high-frequency asymptote of the seed-to-signal NTF caused by the Kerr effect, the reduction of the pump channel caused by the downward shift of its corner frequency, and the decrease of the seed channel, which results from the seed RSD falling more steeply toward high modulation frequencies than the pump RSD. Above approximately 3 W, the frequency shift ceases to be the dominant effect: the seed channel is governed by the Kerr-induced reduction of the NTF asymptote and the pump channel by its corner-frequency shift, so that the numerical RIN falls below the analytical prediction.

Fig. \ref{Fig3}(e) shows the RIN versus fiber length for $f_{\rm rep}=75$ MHz and $P_p=10$ W. The technical RIN remains the dominant contribution, and at this pump power it is governed by the pump channel, as discussed above. Both the numerator and the denominator of Eq. (\ref{xip}) grow with fiber length, the former because more pump power is absorbed and the latter because the output signal power increases. The denominator grows slightly more slowly, however, because it also contains the residual pump power, which decreases with length; consequently, the pump-to-signal NTF at zero modulation frequency increases. In addition, the corner frequency of the pump-to-signal NTF shifts to higher modulation frequencies, which further increases the technical RIN. The numerically calculated RIN remains below the analytical one by approximately the same amount for all fiber lengths, because at this pump power nonlinearities reduce both the seed- and pump-noise contributions, as explained above.

Fig. \ref{Fig3}(f) shows the RIN versus repetition rate for $L_{\rm fiber}=3.5$ m and $P_p=10$ W. Whereas the analytical technical RIN is nearly independent of the repetition rate, the reduction of the numerical RIN by NSB becomes stronger at lower repetition rates. A lower repetition rate implies a longer interval between consecutive pulses, so that each pulse extracts more energy and experiences stronger NSB.

We note that in our previous work \cite{Nitzan} the experimentally measured RIN was compared with the technical RIN obtained in the monochromatic approximation from the pump-to-signal and seed-to-signal NTFs alone, i.e. neglecting the pump-to-ASE and seed-to-ASE contributions. For $L_{\rm fiber} = 2.8\ \rm m$, $P_p = 8\ \rm W$ and $f_{\rm rep} = 75\ \rm MHz$ the measured RIN was $\approx 0.37\,\%$. The monochromatic model without the ASE contributions predicts $0.365\,\%$, whereas the full numerical model predicts $0.34\,\%$. Furthermore, the physical mechanism identified above—wherein nonlinear spectral broadening and gain shaping concentrate the pulse energy in a red-shifted low-noise spectral region while pushing intensity perturbations toward the spectral wings—is consistent with the wavelength-resolved RIN suppression independently demonstrated in an Er-doped gain-dispersion-managed nonlinear amplifier at $1.55 \ \rm \mu m$ \cite{Yang2026}.

\begin{figure} [t]
    \begin{center}
    \hspace*{-1.5 cm}
         \includegraphics[width=140mm]{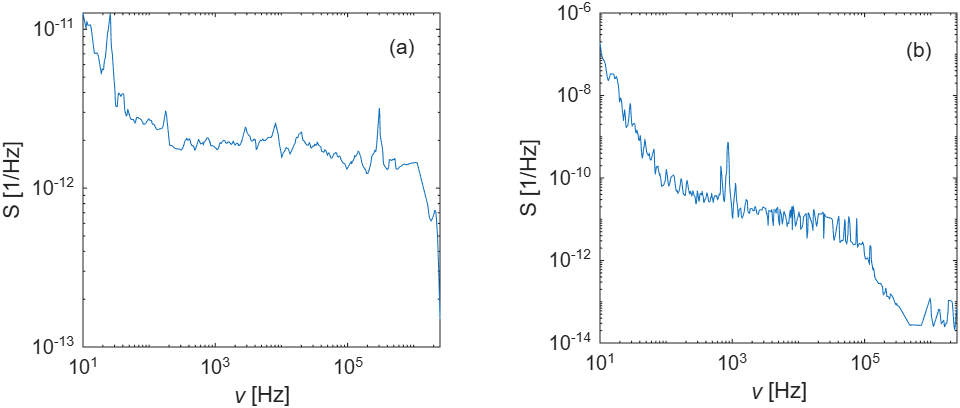}
    \caption{RSDs of the pump (a) and seed (b) for the low-power, low-repetition-rate regime.}
     \label{Fig5}
    \end{center}
\end{figure}

\subsection{Low-power, Low-repetition-rate Nonlinear Regime}
In the second low-power, low-repetition-rate regime, we used a seed pulse with energy $E_s(z=0)=0.05 \ \rm nJ$ and $\mathrm{FWHM}=0.8 \ \rm ps$. The pump power was varied between $0.2$ and $1 \ \rm W$, and the repetition rate between $0.2$ and $5 \ \rm MHz$. In this operating regime, the reduction in the pump power is compensated by the increase in the repetition time, allowing each pulse to accumulate more energy and thereby experience sufficiently strong nonlinear effects for the amplifier to operate in the GMN regime, as demonstrated below. For the RSD of the pump presented in Fig. \ref{Fig5} (a), we used the experimental data given in \cite{Mayer}. For the RSD of the seed depicted in Fig. \ref{Fig5} (b), we used the experimental data given in \cite{Canella}. Importantly, although the RSD of the seed noise remains comparable to that in the high-power, high-repetition-rate regime, the RSD of the pump noise is approximately two orders of magnitude lower over the most of the frequency range. Such a reduction in pump noise is experimentally feasible at the lower pump powers considered here. 

\begin{figure} 
    \begin{center}
    \hspace*{-1.5 cm}
         \includegraphics[width=140mm]{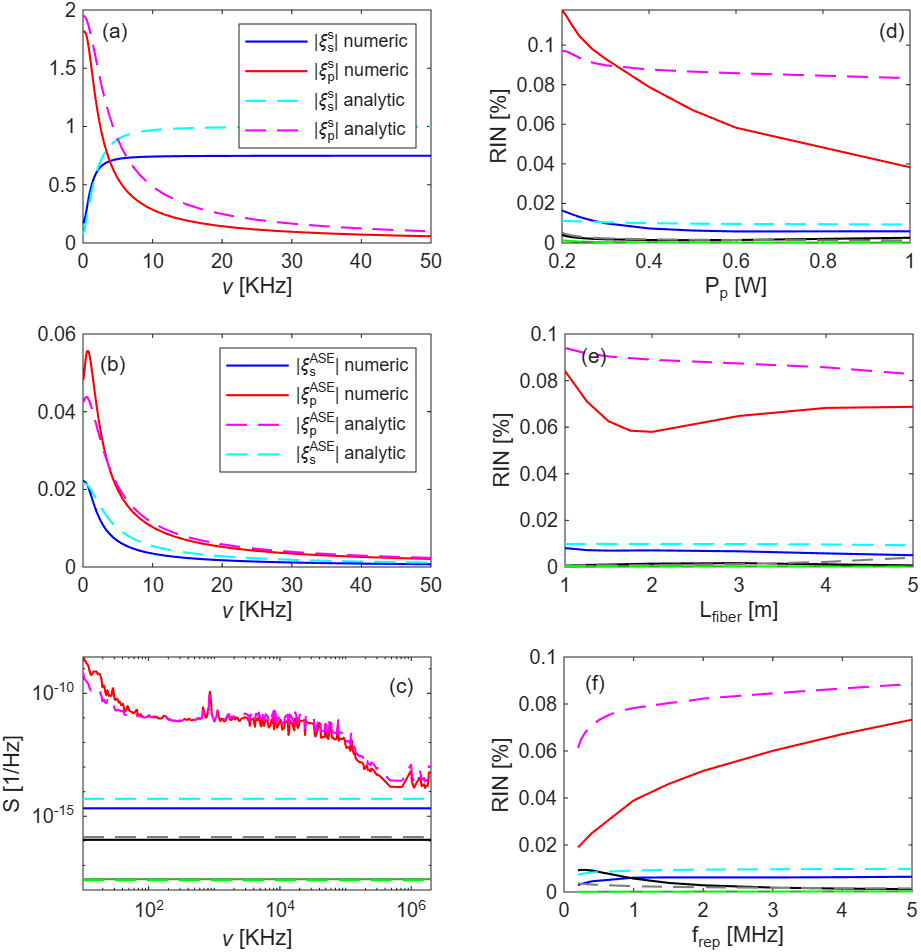}
    \caption{Results for the low-power, low-repetition-rate regime: (a) signal-to-signal and pump-to-signal NTFs, (b) signal-to-ASE and pump-to-ASE NTFs, (c) RSDs of the different noise sources, all plotted as functions of RF for $P_p=0.5 \ W$, $f_{\rm rep}=4 \ \rm MHz$ and $L_{\rm fiber}=3.5 \ \rm m$. In (a-b), numerical pump-to-signal and pump-to-ASE NTFs are shown as solid red curves, whereas the numerical signal-to-signal and signal-to-ASE NTFs are shown as solid blue curves. The analytical pump-to-signal and pump-to-ASE NTFs are shown by dashed magenta curves, and the analytical signal-to-signal and signal-to-ASE NTFs are shown by dashed cyan curves. In (d-f) the RINs of different noise sources are plotted as functions of pump power for $L_{\rm fiber}=3.5 \ \rm m$ and $f_{\rm rep}=4 \ \rm MHz$ (d), as functions of fiber length for $P_p=0.5 \ \rm W$ and $f_{\rm rep}=4 \ \rm MHz$ (e), and as functions of repetition rate for $P_p=0.5 \ \rm W$ and $L_{\rm fiber}=3.5 \ \rm m$ (f). In (c-f), the solid red curves represent numerically calculated results for the technical noise, and the dashed magenta curves represent the corresponding analytical results. The signal–ASE beat noise is shown by the solid blue (numerical) and dashed cyan (analytical) curves. The ASE–ASE beat noise is represented by the solid black (numerical) and dashed gray (analytical) curves. The shot noise is shown by solid dark green (numerical) and dashed light green (analytical) curves.}
     \label{Fig6}
    \end{center}
\end{figure}

\begin{figure} [t]
    \begin{center}
    \hspace*{-1.5 cm}
        \includegraphics[width=140mm]{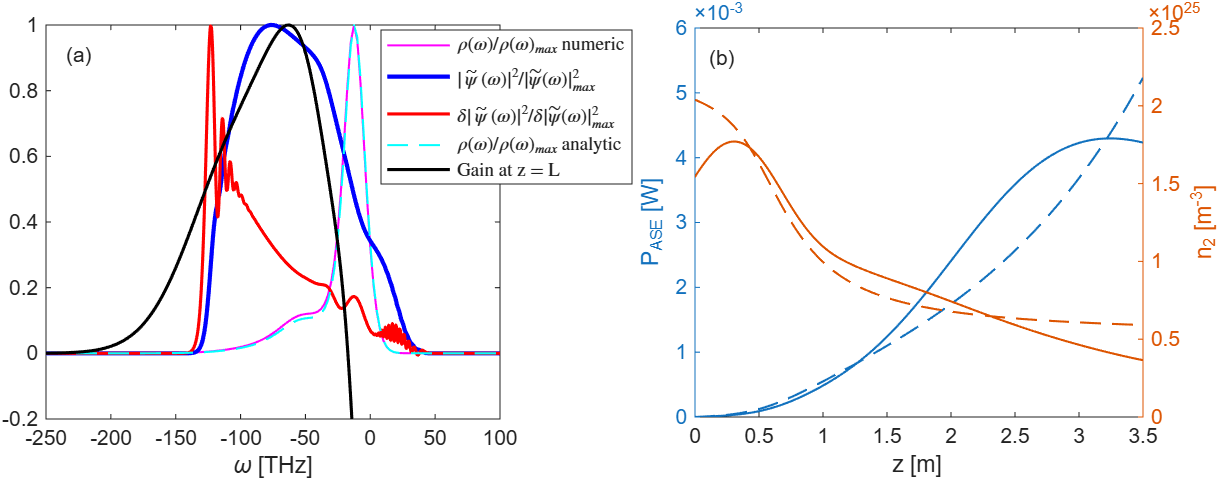}
    \caption{(a) Signal spectrum (solid blue curve) and perturbation of signal (solid red curve) spectrum for $\nu=200 \rm kHz$; Numerical (solid magenta curve) and analytical (dashed cyan curve) ASE PSD. All values are at the fiber output and normalized to their maximum values. (b) Numerical (solid blue curve) and analytical (dashed blue curve) ASE powers, and numerical (solid red curve) and analytical (dashed red curve)  inversion versus propagation length.}
     \label{Fig7}
    \end{center}
\end{figure}

Figure \ref{Fig6}(a) shows the signal-to-signal and pump-to-signal NTFs, Fig. \ref{Fig6}(b) shows the signal-to-ASE and pump-to-ASE NTFs, and Fig. \ref{Fig6}(c) shows the RSDs of the different noise sources as functions of RF for $P_p=0.5 \ \rm W$, $f_{\rm rep}=4 \ \rm MHz$ and $L_{\rm fiber}=3.5 \ \rm m$. 
The curve colors and styles follow the same convention as in Fig. \ref{Fig3}(a-c). 

The qualitative behavior of the NTFs closely resembles that of the high-power, high-repetition-rate regime discussed in the previous subsection. The reduction of numerically calculated high-frequency asymptotic value of signal-to-signal NTF as before can be explained by the Kerr effect that shifts the perturbation spectrum further toward long wavelengths than the signal spectrum, as seen in Fig. \ref{Fig7}(a) which shows the signal spectrum (blue curve), the gain (black curve), and perturbation of the signal spectrum  for $\nu=200 \rm kHz$ (red curve) at the fiber output, each normalized to its maximum value.

The frequency independent numerical and analytical RSD of the ASE–ASE beat noise shown in Fig. \ref{Fig6} (c) are $\approx 1.1 \times 10^{-16} \ {\rm Hz}^{-1}$ and $1.43 \times 10^{-16} \ {\rm Hz}^{-1}$, respectively. As in the previous regime 23 \% reduction of numerical value originates from the lower ASE power in the numerical model, as shown in Fig. \ref{Fig7}(b), which results from the lower inversion in the output section of the fiber, also shown in Fig. \ref{Fig7}(b). The reduced inversion can again be attributed to the nonlinearly broadened signal spectrum, which extends to wavelengths where the absorption cross section is small. The signal therefore continues to extract gain at low inversion and depletes the inversion near the fiber end more strongly than a monochromatic signal. 

The analytically calculated RSD of the signal-ASE beat noise $\approx 5.1 \times 10^{-15} \ {\rm Hz}^{-1}$ is higher than the corresponding numerical value $\approx 2.1 \times 10^{-15} \ {\rm Hz}^{-1}$ by approximately a factor of two and a half, which again is due to the significant red shift and NSB of the signal in GMN regime, resulting in a smaller overlap of the ASE and signal spectra. This is demonstrated in Fig. \ref{Fig7}(a), which shows the numerically calculated ASE PSD (solid magenta curve), the signal spectrum (solid blue curve), and the analytically calculated ASE PSD (dashed cyan curve), all normalized to unity. The RSD of the ASE–ASE beat noise is approximately one order of magnitude lower than that of the signal–ASE. The ASE–ASE beat noise approaches the signal–ASE beat noise because the ASE power is much closer to the signal power in this low-repetition-rate, low-seed-energy regime. The numerically calculated shot-noise PSD ($\approx 2.745 \times 10^{-18} \ {\rm Hz}^{-1}$) is approximately 11 \% higher than the analytical value ($\approx 2.48 \times 10^{-18}\ {\rm Hz}^{-1}$), because the numerical signal power is lower owing to NSB of the signal in combination with the frequency-dependent gain.  

Figures \ref{Fig6}(d-f) show the integrated relative intensity noises (RINs) associated with the different noise sources. These RINs were evaluated using Eq. (\ref{RINtech}) with corresponding RSDs, and $\nu_{\rm min}=10 \ {\rm Hz}$, $\nu_{\rm max}=f_{\rm rep}/2$. The curve colors and styles follow the same convention as in Fig. \ref{Fig3}(d-f). 

Figure \ref{Fig6}(d) depicts RINs versus pump power for $f_{\rm rep}=4 \ \rm MHz$, $L_{\rm fiber}=3.5 \ \rm m$. Technical noise remains dominant in this regime, although the ASE–ASE and signal–ASE beat noise contributions become comparable to it. Due the significantly lower pump RSD RIN is dominated by seed contribution. Increasing the pump power shifts the corner frequency of the signal-to-signal NTF toward higher frequencies, thereby reducing the analytically calculated technical RIN. The numerically calculated technical RIN decreases considerably faster with pump power than its analytical counterpart. This behavior can be understood as follows. As explained above, the asymmetry of the gain converts the symmetric NSB into a red shift that grows with pulse energy. In the low-pump-power, low-repetition-rate regime the pulse energies are small, and for $P_p \lesssim 0.3 \rm W$ the output spectrum still lies on the short-wavelength slope of the gain curve. An increase in pulse energy then shifts the spectrum toward higher gain, so that the output energy grows faster than the input energy, and the high-frequency asymptote of the seed-to-signal NTF exceeds unity. This mechanism accounts for the numerical technical RIN exceeding its analytical counterpart at $P_p \lesssim 0.3 \rm W$. At $P_p \gtrsim 0.3 \rm W$ the pulse acquires more energy and its spectrum shifts further toward longer wavelengths, onto the long-wavelength slope of the gain curve. An increase in pulse energy then moves the spectrum toward lower gain, so that the Kerr effect reduces the high-frequency asymptote of the seed-to-signal NTF below unity. This effect accounts for the more rapid decrease of the numerical RIN with pump power compared with the analytical prediction.

Figure \ref{Fig6}(e) shows the RIN as a function of fiber length for $f_{\rm rep}=4 \ \rm MHz$ and $P_p=0.5 \ \rm W$. The analytical RIN decreases with fiber length because the increasing signal power shifts the corner frequency of the signal-to-signal NTF toward higher frequencies and in addition reduces the DC value of the NTF as explained above.

The non-monotonic behavior of the numerical technical RIN can be attributed to the Kerr-induced variation of the high-frequency asymptotic values of the signal-to-signal NTF along the fiber. During the first $2 \ \rm m$  of propagation, the population inversion is relatively high, resulting in a gain spectrum peaked near $\lambda=1030 \ \rm nm$ . In this regime, stronger NSB shifts spectrum further toward longer wavelengths, onto the long-wavelength slope of the gain curve thereby reducing the effective gain, the asymptotic values of the signal-to-signal NTF, and consequently the technical RIN. Beyond $2 \ \rm m$, the lower population inversion shifts the gain peak toward longer wavelengths, so that it approaches the centroid of the red-shifted signal spectrum. Under these conditions the red shift caused by NSB is penalised much less than at $2\ \rm m$, where the spectrum already lies on the long-wavelength slope of the gain. Consequently, the asymptotic value of the seed-to-signal NTF partially recovers, reaching $0.72$ at $3 \ \rm m$ and $0.76$ at $5 \ \rm  m$, which in turn increases the numerical technical RIN.

Figure \ref{Fig6}(f) shows the RIN as a function of repetition rate for $L_{\rm fiber}=3.5 \ \rm m$ and $P_p=0.5 \ \rm W$. Repetition rate varies from $200 \rm kHz$ to $5 \rm MHz$. The technical RIN decreases with decreasing repetition rate in both models, but for a reason unrelated to the gain dynamics. The corner frequency and the DC value of the seed-to-signal NTF are governed by the average signal and pump powers, which vary only weakly with repetition rate. Instead, the decrease originates from the growth of ASE: as the repetition rate is reduced, the average seed power decreases, the signal extracts a smaller share of the absorbed pump power, and the remaining stored energy is released as ASE. The resulting reduction of the factor $P_s/(P_s+P_{\rm ASE})$ in Eq. (\ref{RINtech}) lowers the technical RIN, and the truncation of the integration band at $f_{\rm rep}/2$ removes the seed noise above this frequency. The numerical RIN falls considerably faster because the full model accounts for gain saturation by ASE, which the analytical model neglects. At $f_{\rm rep}=250$ kHz the ASE carries $62 \% $  of the output power in the numerical model but only $23 \%$ in the analytical one. The Kerr-induced reduction of the NTF asymptote, which halves the seed channel at low repetition rates, is compensated by the growth of the pump channel, since the signal experiences a high unsaturated gain and becomes highly sensitive to inversion fluctuations. Under these conditions the RIN contributions of the ASE–ASE and signal–ASE beat noises become comparable to the technical RIN: at $f_{\rm rep}=250$ kHz, $\rm RIN_{tech}=0.02\,\%$, $\rm RIN_{ASE-ASE}=0.0094\,\%$ and $\rm RIN_{signal-ASE}=0.0034\,\%$.

\section{Conclusion}
In conclusion, we have developed a rigorous theoretical framework for predicting pulse-to-pulse power fluctuations caused by the dominant noise sources, namely pump and seed power fluctuations, ASE-ASE and signal-ASE beat noises and shot noise, in ultrafast fiber amplifiers operating in the GMN regime, and have investigated the influence of NSB on these fluctuations. The numerical model evaluates statistically averaged quantities directly, avoiding computationally intensive Monte Carlo simulations with stochastic source terms. Its predictions were compared with those of a semi-analytical monochromatic model that neglects both the NSB of the signal and the depletion of the inversion by ASE; in the limit of vanishing signal bandwidth and negligible ASE the two models were shown to coincide. In the GMN regime, the NSB of the signal and of its perturbations, acting together with the frequency-dependent gain, significantly alters the technical RIN in both the high-power, high-repetition-rate and the low-power, low-repetition-rate regimes. In the latter regime ($\sim 0.5$ W, $\sim 0.25$ MHz) the RIN contributions of the ASE-related noise sources become comparable to those induced by pump and seed fluctuations. To the best of our knowledge, this work also provides the first analytical expressions for the signal- and pump-to-ASE noise transfer functions, together with improved expressions for the signal–ASE and ASE–ASE beat-noise intensities. 

Finally, we note that the results above correspond to the total unfiltered output at the fiber exit, where the photodetector collects both the broadband signal and the entire ASE spectrum. In many practical implementations, the amplifier is followed by a grating compressor or a long-pass spectral filter. Within our framework, any passive post-amplifier filter with transmission $T_{\rm filt} (\omega)$ can be incorporated directly at $z=L$ —without re-solving the propagation Eqs. (\ref{GNLSE} - \ref{drof}) - by multiplying the output signal and ASE spectra (and their perturbations) by $T_{\rm filt} (\omega)$ in the detector integrals [Eqs. (\ref{SE}–\ref{dro_f}), (\ref{P0ASE} -- \ref{ASE-ASE})]. In particular, because the GMN signal is strongly red-shifted relative to the ASE peak [see Figs. \ref{Fig4}(a) and \ref{Fig7}(a)], a long-pass filter or compressor aperture that rejects the short-wavelength ASE band will strongly suppress the ASE–ASE beat noise and the signal/pump-to-ASE transfer functions while restoring the power-normalization factor $P_{s}^0/(P_{s}^0 + P_{ASE}^0)$ in Eq. (\ref{RINtech}) toward unity, whereas the signal–ASE beat noise [Eq. (\ref{Signal-ASE})] remains governed by the in-band ASE overlapping the red-shifted signal. Moreover, since the signal energy perturbation is concentrated at the outer spectral wings [red curves in Figs. \ref{Fig4}(a) and \ref{Fig7}(a)], any spectral filtering on the long-wavelength edge of the pulse will further reduce the seed-to-signal NTF.

\section*{Appendix A}
In this appendix, we derive generalized formulas for the Signal-ASE and ASE-ASE beat noises, for arbitrary spectral properties of the signal and ASE. The signal field is given by
\begin{equation}
    E_s\bigl(t) =\sqrt{\frac{n}{2 \pi}}\int d \omega |\tilde{\psi}\left( \omega \right)|e^{-i \left[\omega t-\Phi\left(\omega\right)\right]}\label{Signal} ,
\end{equation}
where $n$ is the number of pulses. Strictly speaking, the following analysis is exact only in the limit when $n \to \infty$, however, we will omit $\lim_{n \to \infty}$ in what follows for brevity.
The ASE field
\begin{equation}
    E_{\rm ASE}\bigl(t) =\sum_k\sqrt{\rho \bigl( \omega_k \bigr)\Delta \omega } e^{-i \left[\omega_k t-\phi\left(\omega_k\right)\right]}\label{ASE_D} ,
\end{equation}
where $\Delta \omega=2 \pi /\left(n T_{\rm rep}\right)$ and $ \omega_k=k \Delta \omega$. In the limit when $n \to \infty$, one finds that
\begin{equation}
    E_{\rm ASE}\bigl(t) =\sqrt{\frac{T_{\rm rep}n}{2 \pi}} \int d \omega \sqrt{\rho \bigl( \omega \bigr)} e^{-i \left[\omega t-\phi\left(\omega\right)\right]}\label{ASE}.
\end{equation}
The low-frequency Signal-ASE beat intensity is given by
\begin{equation}
\begin{split}
I&=E_s E_{\rm ASE}^*+E_s^* E_{\rm ASE}  \\
& =\frac{2 n \sqrt{T_{\rm rep}}}{ 2 \pi} \int d \omega \int d \omega' |\tilde{\psi}\left( \omega \right)| \sqrt{\rho\left( \omega' \right)} \cos\left[ (\omega-\omega')t + \Phi\left( \omega \right) - \phi\left(  \omega'\right)\right],
\end{split}
\end{equation}

from which it follows that
\begin{equation}
\begin{split}
\left< I^2 \right> & =\frac{4 n^2  T_{\rm rep}}{ \left(2 \pi\right)^2}  \\
&  \times \left< \{ \int d \omega \int d \omega' |\tilde{\psi}\left( \omega \right)| \sqrt{\rho\left( \omega' \right)} \cos\left[ (\omega-\omega')t + 
\Phi\left( \omega \right) - \phi\left(  \omega'\right)\right] \} ^2 \right>.
\end{split}
\end{equation}
By neglecting the double frequency terms, one finds that
\begin{equation}
\begin{split}
\left< I^2 \right> & =\frac{2 n^2  T_{\rm rep}}{ \left(2 \pi\right)^2}  \int d \omega \int d \omega' \int d \omega'' \int d \omega''' \\
& \times |\tilde{\psi}\left( \omega \right)| |\tilde{\psi}\left( \omega'' \right)| \sqrt{\rho\left( \omega' \right)} \sqrt{\rho\left( \omega''' \right)}e^{i \left[ \Phi\left( \omega \right)-\Phi\left( \omega'' \right)\right]} \\
& \times \left< e^{i \left[ \left(\omega-\omega' - \omega''+\omega'''\right)t+ \phi\left(  \omega'''\right)-\phi\left(  \omega'\right)\right]} \right> .
\end{split} \label{11}
\end{equation}
Because of the random nature of ASE we can factorize 
\begin{equation}
\begin{split}
& \left< e^{i \left[ \left(\omega-\omega' - \omega''+\omega'''\right)t+ \phi\left(  \omega'''\right)-\phi\left(  \omega'\right)\right]} \right> \\
& =\left< e^{i \left[ \omega-\omega'' - \omega'+\omega'''\right]t} \right>\left< e^{i\left[\phi\left( \omega''' \right) - \phi\left(  \omega'\right)\right]} \right> \\
& =\left( \frac{2 \pi}{n T_{\rm rep}} \right)^2 \delta\left( \omega-\omega'' - \omega'+\omega''' \right) \delta\left( \omega'-\omega''' \right) \\
& =\left( \frac{2 \pi}{n T_{\rm rep}} \right)^2 \delta\left( \omega-\omega'' \right) \delta\left( \omega'-\omega''' \right)  .
\end{split} \label{21}
\end{equation}

By substituting Eq. (\ref{21}) into Eq. (\ref{11}), one obtains
\begin{equation}
    \left< I^2 \right>  =\frac{2 }{  T_{\rm rep} } \int d \omega \int d \Delta \omega  |\tilde{\psi}\left( \omega \right)| ^2 \rho\left( \omega + \Delta \omega \right).
\end{equation}
The RSD of the signal-ASE beat noise is given by 
\begin{equation}
    S\left( \Delta \omega \right) = \frac{1}{2 P^2_s} \frac{d\left< I^2 \right>}{d \Delta \omega}, \label{rsd_p}
\end{equation}
where the factor of $1/2$ comes from polarization and $P_s=\int d \omega  |\tilde{\psi}\left( \omega \right)| ^2 /T_{\rm rep} $.
Then the one-sided RSD of the signal-ASE beat noise is
\begin{equation}
    S_{\rm S-ASE}\left( \Delta \omega\right) = T_{\rm rep} \frac{\int d \omega  |\tilde{\psi}\left( \omega \right)| ^2 \left[ \rho \left( \omega + \Delta \omega\right) + \rho \left( \omega - \Delta \omega\right)\right]}{\left[ \int d \omega  |\tilde{\psi}\left( \omega \right)| ^2\right]^2},
\end{equation}
or, in terms of the radio-frequency $\nu$
\begin{equation}
    S_{\rm S-ASE}\left( \nu \right) = 2 \pi T_{\rm rep} \frac{\int d \omega  |\tilde{\psi}\left( \omega \right)| ^2 \left[ \rho \left( \omega + 2 \pi \nu \right) + \rho \left( \omega - 2 \pi \nu \right)\right]}{\left[ \int d \omega  |\tilde{\psi}\left( \omega \right)| ^2\right]^2}.
\end{equation}
The low-frequency ASE-ASE beat intensity is given by
\begin{equation}
\begin{split}
I&=E_{\rm ASE}^* E_{\rm ASE}  \\
& =\frac{n T_{\rm rep}}{ 2 \pi} \int d \omega \int d \omega' \sqrt{\rho\left( \omega \right)} \sqrt{\rho\left( \omega' \right)} e^{i \left[ (\omega-\omega')t + \phi\left( \omega \right) - \phi\left(  \omega'\right)\right]}.
\end{split}
\end{equation}
By neglecting the double frequency terms, one finds that
\begin{equation}
\begin{split}
\left< I^2 \right> & =\left(\frac{n  T_{\rm rep}}{ 2 \pi} \right)^2 \int d \omega \int d \omega' \int d \omega'' \int d \omega''' \\
& \times \sqrt{\rho\left( \omega \right)} \sqrt{\rho\left( \omega'' \right)}  \sqrt{\rho\left( \omega' \right)} \sqrt{\rho\left( \omega''' \right)} \\
& \times \left< e^{i \left[ \left( \omega-\omega' - \omega''+\omega''' \right)t-\phi\left( \omega'' \right)+\phi\left( \omega \right)-\phi\left(  \omega'\right)+ \phi\left(  \omega'''\right)\right]} \right> .
\end{split} \label{3}
\end{equation}
Since we are interested in the non-DC photocurrent, $\omega \ne \omega'$ and so the random phases $\phi\left( \omega \right)$ and $\phi\left( \omega' \right)$ are uncorrelated and we can factorize 
\begin{equation}
\begin{split}
& \left< e^{i \left[ \left( \omega-\omega' - \omega''+\omega''' \right)t-\phi\left( \omega'' \right)+\phi\left( \omega \right)-\phi\left(  \omega'\right)+ \phi\left(  \omega'''\right)\right]} \right>\\
& =\left< e^{i \left[ \left( \omega-\omega'' \right)t+\phi\left( \omega \right) - \phi\left(  \omega''\right)\right]} \right>\left< e^{i \left[ \left( - \omega'+\omega'''\right)t-\phi\left( \omega' \right) + \phi\left(  \omega'''\right)\right]} \right> \\
& =\left( \frac{2 \pi}{n T_{\rm rep}} \right)^2 \delta\left( \omega-\omega'' \right) \delta\left( \omega'-\omega''' \right).
\end{split} \label{31}
\end{equation}

By substituting Eq. (\ref{31}) into Eq. (\ref{3}), one finds that
\begin{equation}
   \left< I^2 \right>=\int d \omega \int d \omega' \rho\left( \omega \right)\rho\left( \omega' \right).
\end{equation}
The one-sided RSD of the ASE-ASE beat noise is given by
 \begin{equation}
    S_{\rm ASE-ASE}\left( \Delta \omega\right) = \frac{\int d \omega  \rho\left( \omega \right)\rho\left( \omega + \Delta \omega \right)}{P^2_s},
\end{equation}
where the factor of $1/2$ is due to polarization
or, in terms of radio-frequency $\nu$,
\begin{equation}
    S_{\rm ASE-ASE}\left( \nu \right) = 2 \pi \frac{\int d \omega  \rho\left( \omega \right)\rho\left( \omega + 2 \pi \nu \right)}{P^2_s}.
\end{equation}

\section*{Appendix B}
In this appendix, we present an alternative derivation of Eq. (\ref{K_ASE}). By using Eq. (\ref{n2q}), it is easy to show that 
\begin{equation}
   n_2^0(\zeta) q_n(\zeta) = \frac{1}{A_d}\frac{d}{d \zeta} \frac{X_k(\zeta)}{D(\zeta)}, \label {nn2q}
\end{equation}
where the subscript $k=p$ for the initial conditions [$q_p(\nu,0)=1, q_s(\nu,0)=0$] and $k=s$ for the initial conditions [$q_s(\nu,0)=1, q_p(\nu,0)=0$].
By substituting Eq. (\ref{nn2q}) into Eq. (\ref{drof}) one finds that
\begin{equation}
\begin{split}
    d_z \delta \rho (\omega, \zeta) & = \bigl[\left(\beta_{\omega} + \mu_{\omega}\right) n^0_2 -\mu_{\omega} n_{\rm tot} \bigr] \delta \rho (\omega, \zeta) \\
    & +\left[ 2 \beta_{\omega} \hbar \left( \omega_s+\omega \right)/\left( 2 \pi \right) + \left(\beta_{\omega} + \mu_{\omega}\right) \rho_0 \right] \frac{1}{A_d}\frac{d}{d \zeta} \frac{X_k(\zeta)}{D(\zeta)}. \label{roLin3}    
\end{split}
\end{equation}
The solution of Eq. (\ref{roLin3}) is given by
\begin{equation}
\begin{split}  
 \delta \rho (\omega, z) & = \int_0^z d \zeta \left[ \frac{2 \beta_{\omega} \hbar \left( \omega_s+\omega \right)}{\left( 2 \pi \right)} + \left(\beta_{\omega} + \mu_{\omega}\right) \rho_0 \right] \frac{1}{A_d}\frac{d}{d \zeta} \frac{X_k(\zeta)}{D(\zeta)} \\
  & \times \exp{\bigl[\bigl(\beta_{\omega} + \mu_{\omega}\bigr)\frac{N_2^0(z)-N_2^0(\zeta)}{A_d} - \mu_{\omega} n_{\rm tot}\bigl(z-\zeta\bigr) \bigr]}, \label{drho}
\end{split}
\end{equation}
By substituting Eq. (\ref{ros0}) for $\rho_0$ into Eq. (\ref{drho}), one finds that 
\begin{equation}
\begin{split}  
  \delta \rho (\omega, z) & = \frac{2 \beta_{\omega}   \hbar \left( \omega_s+\omega \right)}{\left( 2 \pi \right)} \int_0^z d \zeta  \frac{1}{A_d}\frac{d}{d \zeta} \frac{X_k(\zeta)}{D(\zeta)} \\
  & \times \biggl\{1 + \left(\beta_{\omega} + \mu_{\omega}\right) \int_0^\zeta d \xi n_2^0 (\xi) \\ & \times \exp{\bigl[\bigl(\beta_{\omega} + \mu_{\omega}\bigr)\frac{N_2^0(\zeta)-N_2^0(\xi)}{A_d} - \mu_{\omega} n_{\rm tot}\bigl(\zeta-\xi\bigr) \bigr]} \biggr\}  \\
  & \times \exp{\bigl[\bigl(\beta_{\omega} + \mu_{\omega}\bigr)\frac{N_2^0(z)-N_2^0(\zeta)}{A_d} - \mu_{\omega} n_{\rm tot}\bigl(z-\zeta\bigr) \bigr]} \\
  & = \frac{2 \beta_{\omega}   \hbar \left( \omega_s+\omega \right)}{\left( 2 \pi \right)} \int_0^z d \zeta  \frac{1}{A_d}\frac{d}{d \zeta} \frac{X_k(\zeta)}{D(\zeta)} \\
  & \times \biggl\{  \exp{\bigl[\bigl(\beta_{\omega} + \mu_{\omega}\bigr)\frac{N_2^0(z)-N_2^0(\zeta)}{A_d} - \mu_{\omega} n_{\rm tot}\bigl(z-\zeta\bigr) \bigr]}
   + \left(\beta_{\omega} + \mu_{\omega}\right) \\ & \times \int_0^\zeta d \xi n_2^0 (\xi)  \exp{\bigl[\bigl(\beta_{\omega} + \mu_{\omega}\bigr)\frac{N_2^0(z)-N_2^0(\xi)}{A_d} - \mu_{\omega} n_{\rm tot}\bigl(z-\xi\bigr) \bigr]} \biggr\}. \label{ros3}
\end{split}
\end{equation}
By integrating by parts the second term in the curly brackets, one finds that
\begin{equation}
    \begin{split}
        & \frac{2 \beta_{\omega}   \hbar \left( \omega_s+\omega \right)}{\left( 2 \pi \right)} \int_0^z d \zeta \frac{1}{A_d}\frac{d}{d \zeta} \frac{X_k(\zeta)}{D(\zeta)} \\
  & \times  \left(\beta_{\omega} + \mu_{\omega}\right) \int_0^\zeta d \xi n_2^0 (\xi)  \exp{\bigl[\bigl(\beta_{\omega} + \mu_{\omega}\bigr)\frac{N_2^0(z)-N_2^0(\xi)}{A_d} - \mu_{\omega} n_{\rm tot}\bigl(z-\xi\bigr) \bigr]}   \\
  & = \frac{2 \beta_{\omega}   \hbar \left( \omega_s+\omega \right)}{\left( 2 \pi \right)} \int_0^z d \zeta n_2^0 (\zeta) \frac{\beta_{\omega}+\mu_{\omega}}{ A_d} \biggl(\frac{X_k(z)}{D(z)} - \frac{X_k(\zeta)}{D(\zeta)} \biggr) \\
  & \times \exp{\bigl[\bigl(\beta_{\omega} + \mu_{\omega}\bigr)\frac{N_2^0(z)-N_2^0(\zeta)}{A_d} - \mu_{\omega} n_{\rm tot}\bigl(z-\zeta\bigr) \bigr]} , \label{rosd3}
\end{split}
\end{equation}
$X_k(0)=0$ was used. Subsequently, Eq. (\ref{ros3}) becomes
\begin{equation}
\begin{split}
  \delta \rho (\omega, z) & =\frac{2 \beta_{\omega} \hbar (\omega_s+\omega)}{(2 \pi)} \\ & \times \int_0^z d \zeta 
  \biggl[ n_2^0 (\zeta) \frac{\beta_{\omega}+\mu_{\omega}}{ A_d} \biggl(\frac{X_k(z)}{D(z)} - \frac{X_k(\zeta)}{D(\zeta)} \biggr)   + \frac{1}{A_d}\frac{d}{d \zeta} \frac{X_k(\zeta)}{D(\zeta)}\biggr] \\
 & \times \exp{\bigl[\bigl(\beta_{\omega} + \mu_{\omega}\bigr)\frac{N_2^0(z)-N_2^0(\zeta)}{A_d} - \mu_{\omega} n_{\rm tot}\bigl(z-\zeta\bigr) \bigr]}   \label{ro_ASE}
\end{split}
\end{equation}


\newpage

\begin{thebibliography} {1}
\newcommand{\enquote}[1]{``#1''}

\bibitem{Ferm2}	
M.~E. Fermann and I.~Hartl,
\enquote{Ultrafast fiber laser technology,}  {\protect\JournalTitle{IEEE J. Sel. Top. Quantum Electron}} \textbf{15}, 191--206 (2009).

\bibitem{Ferm1}	
M.~E. Fermann and I.~Hartl,
\enquote{Ultrafast fibre lasers,}  {\protect\JournalTitle{Nat. Photonics}} \textbf{7}, 868--874 (2013). 

\bibitem{Zhao}	
W.~Zhao, X.~Hu, and Y.~Wang,
\enquote{Femtosecond-pulse fiber based amplification techniques and their applications,}
{\protect\JournalTitle{IEEE J. Sel. Top. Quantum Electron}} \textbf{20}, 512--524  (2014). 

\bibitem{Chang}	
G.~Chang, and Z.~Wei,
\enquote{Ultrafast fiber lasers: an expanding versatile toolbox,}
{\protect\JournalTitle{iScience}} \textbf{23}, 101101  (2020).

\bibitem{Perry}
M.~D. Perry, T. Ditmire, and B.~C. Stuart, 
\enquote{Self-phase modulation in chirped-pulse amplification,} 
{\protect\JournalTitle{Optics Letters}} \textbf{19}, 2149--2151 (1994).

\bibitem{Morou}
D.~Strickland and G.~Mourou,
\enquote{Compression of amplified chirped optical pulses,}  {\protect\JournalTitle{Optics Communications}} \textbf{56}, 219--221 (1985).


\bibitem{Shank}
C.~Shank, R.~Fork, R.~Yen, R.~Stolen, W.~Tomlinson,
\enquote{Compression of femtosecond optical pulses,}  {\protect\JournalTitle{Applied Physics Letters}} \textbf{40}, 761--763 (1982).

\bibitem{Zaout}
 Y.~Zaouter, D.~N. Papadopoulos, M.~Hanna, J.~Boullet, L.~Huang,
C.~Aguergaray, F.~Druon, E.~Mottay, P.~Georges, and E.~Cormier,
\enquote{Stretcher-free high energy nonlinear amplification of femtosecond pulses in rod-type fibers,}  {\protect\JournalTitle{Optics Letters}} \textbf{33}, 107--109 (2008).

\bibitem{Liu1}
W.~Liu, D.~N. Schimpf, T.~Eidam, J.~Limpert, A.~Tünnermann, F.~X.
Kärtner, and G.~Chang,
\enquote{Pre-chirp managed nonlinear amplification in
fibers delivering 100 W, 60 fs pulses,}  {\protect\JournalTitle{Optics Letters}} \textbf{40}, 151--154 (2015).

\bibitem{Chen}
H.~W. Chen, J. Lim, S.~W. Huang, D.~N. Schimpf, F.~X. Kärtner, and G.~
Chang,
\enquote{ Optimization of femtosecond Yb-doped fiber amplifiers for high-quality pulse compression,}  {\protect\JournalTitle{Optics Express}} \textbf{20}, 28672--28682 (2012).

\bibitem{And} 
D.~Anderson, M.~Desaix, M.~Karlsson, M.~Lisak, and M.~L. Quiroga-Teixeiro,
\enquote{Wave-breaking-free pulses in nonlinear-optical fibers,}  {\protect\JournalTitle{Journal of Optical Society of America B}} \textbf{10}, 1185--1190 (1993).

\bibitem{Krug}
V.~I. Kruglov, A.~C. Peacock, J.~M. Dudley, and J.~D. Harvey,
\enquote{Self-similar propagation of high-power parabolic pulses in optical fiber amplifiers,}  {\protect\JournalTitle{Optics Letters}} \textbf{25}, 1753--1755 (2000).

\bibitem{Ferm} 
M.~E. Fermann, V.~I. Kruglov, B.~C. Thomsen, J.~M. Dudley, and J.~D.
Harvey,
\enquote{Self-similar propagation and amplification of parabolic pulses
in optical fibers,}  {\protect\JournalTitle{Physical Review Letters}} \textbf{84}, 6010--6013 (2000).

\bibitem{Liu2}
Y.~Liu, W.~Li, D.~Luo, D.~Bai, C.~Wang, and H.~Zeng,
\enquote{Generation of 33 fs
935 W average power pulses from a third-order dispersion managed
self-similar fiber amplifier,}  {\protect\JournalTitle{Optics Express}} \textbf{24}, 10939--10945 (2016).

\bibitem{Song}
H.~Song, B.~Liu, Y.~Li, Y.~Song, H.~He, L.~Chai, M.~Hu, and C.~Wang,
\enquote{Practical 24-fs, 1-$\mu$J, 1-MHz Yb-fiber laser amplification system,}  {\protect\JournalTitle{Optics Express}} 
\textbf{25}, 7559--7566 (2017).

\bibitem{Sidor}	
P.~Sidorenko, W.~Fu, and F.~Wise,
\enquote{Nonlinear ultrafast fiber amplifiers beyond the gain-narrowing limit,}  {\protect\JournalTitle{Optica}} \textbf{6}, 1328--1333 (2019).

\bibitem{Sidor2}
P.~Sidorenko, and F.~Wise,
\enquote{Generation of 1 $\mu$J and 40 fs pulses from a large mode area gain-managed nonlinear amplifier,}  {\protect\JournalTitle{Optics Letters}} \textbf{45}, 4084--4087 (2020).

\bibitem{Xia}
T.~Xia, H.~Zhao , Z.~Cheng, Y.~Hou, Q.~Zhang, S.~Cao, Y.~Chen, P.~Zhang, and P.~Wang,
\enquote{All-fiber integrated gain-managed nonlinear amplification system delivers ultrafast lasers with 3.6 MW peak power and 45 fs pulse duration,} 
{\protect\JournalTitle{Optics and Laser Technology}} \textbf{181}, 111627 (2025).

\bibitem{Stol}
D.~Stoliarov, E.~Manuylovich, A.~Koviarov, D.~Galiakhmetova, and E.~Rafailov,
\enquote{Gain-managed nonlinear amplification of
 ultra-long mode-locked fiber laser,} 
{\protect\JournalTitle{Optics Express}} \textbf{31}, 
43427--43437 (2023).

\bibitem{Tom}
D.~Tomaszewska‑Rolla, R.~Lindberg, V.~Pasiskevicius, F.~Laurell, and G.~Soboń,
\enquote{A comparative study of an Yb‑doped fiber gain‑managed 
nonlinear amplifier seeded by femtosecond fiber lasers,}  {\protect\JournalTitle{Sci. Rep.}} \textbf{12}, 404 (2022).

\bibitem{Mik}
M.~Krakowski, and G.~Soboń,
\enquote{Gain-managed nonlinear amplification in an erbium-doped fiber,} 
{\protect\JournalTitle{Optics Express}} \textbf{32}, 
48815-48823 (2024).

\bibitem{Yang2026} M. Yang, Y. Chen, T. Liu, Z. Wan, Y. Dai, S. Xu, Z. Guo, K. Huang, M. Yan, and H. Zeng, \enquote{Noise suppression via gain-dispersion-managed nonlinear amplification in a high-power Er-fiber laser,} {\protect\JournalTitle{Laser Photonics Rev.}} \textbf{2026} , e71831 (2026).

\bibitem{Zhao2017} J. Zhao, G. Guiraud, F. Floissat, B. Gouhier, S. Rota-Rodrigo, N. Traynor, and G. Santarelli, \enquote{Gain dynamics of clad-pumped Yb-fiber amplifier and intensity noise control,} Opt. Express \textbf{25}, 357–366 (2017).

\bibitem{Gierschke2019} P. Gierschke, C. Jauregui, T. Gottschall, and J. Limpert, \enquote{Relative amplitude noise transfer function of an $\rm Yb^{3+}$-doped fiber amplifier chain,} {\protect\JournalTitle{Opt. Express}} \textbf{27}, 17041–17050 (2019).

\bibitem{Han2021} J. Han, Y. Liu, Z. Deng, L. Zhou, C. Li, Z. Luo, and W. Li, \enquote{Optimizing noise characteristics of mode-locked Yb-doped fiber laser using gain-induced RIN-transfer dynamics,} {\protect\JournalTitle{High Power Laser Sci. Eng.}} \textbf{9}, e36 (2021).

\bibitem{Kim2016} J. Kim and Y. Song, \enquote{Ultralow-noise mode-locked fiber lasers and frequency combs: principles, status, and applications,} {\protect\JournalTitle{Adv. Opt. Photon.}} \textbf{8}, 465–540 (2016).

\bibitem{Did} L. Nugent-Glandorf, T. A. Johnson, Y. Kobayashi, and S. A. Diddams, \enquote{Impact of dispersion on amplitude and frequency
noise in a Yb-fiber laser comb,} {\protect\JournalTitle{Optics Letters}} \textbf{36}, 1578-1580 (2011).

\bibitem{Roh} F. Rohde, E. Benkler, and H. R. Telle, \enquote{High contrast, low noise selection and amplification of an individual optical frequency comb line,} {\protect\JournalTitle{Optics Letters}} \textbf{38}, 103-105 (2013).

\bibitem{Bra}
A.D. Brandt, S.F. Cooper, Z. Burkley, D.C. Yost,
\enquote{Reduced phase 
noise in an erbium frequency comb via intensity noise suppression,} {\protect\JournalTitle{Optics Express}} \textbf{25}, 
18175–18181 (2017).


\bibitem{bode}
N. Bode, J. Briggs, X. Chen, M. Frede, P. Fritschel, M. Fyffe,
E. Gustafson, M. Heintze, P. King, J. Liu, J. Oberling,
R. L. Savage, A. Spencer, and B. Willke,
\enquote{Advanced LIGO Laser Systems for O3 and Future
Observation Runs,} 
{\protect\JournalTitle{Galaxies}} \textbf{84}, 
8 (2020).

\bibitem{Cah}
C. Cahillane, G. L. Mansell, and D. Sigg,
\enquote{Laser frequency noise in next generation gravitational-wave detectors,} 
{\protect\JournalTitle{Optics Express}} \textbf{29}, 
42144-42161 (2021).

\bibitem{Kwee} P. Kwee, B. Willke, and K. Danzmann, \enquote{Shot-noise-limited laser power stabilization with a high-power photodiode array,} {\protect\JournalTitle{Optics Letters}} \textbf{34}, 2912-2914 (2009).

\bibitem{Buik} A. Buikema, C. Cahillane, G. L. Mansell, C. D. Blair, R. Abbott, C. Adams, R. X. Adhikari, A. Ananyeva, and S. Appert, \enquote{Sensitivity and performance of the Advanced LIGO detectors in the third observing run,} {\protect\JournalTitle{Phys. Rev. D}} \textbf{102}, 062003 (2020).

\bibitem{Udem} T. Udem, R. Holzwarth, T.W. Hänsch, \enquote{Optical frequency metrology,} {\protect\JournalTitle{Nature}} \textbf{416}, 233 (2002).

\bibitem{Nitzan} N. Haviv, M. Kozlov, M. Kruger, and P. Sidorenko, \enquote{Noise suppression in gain-managed nonlinear
amplifiers,} {\protect\JournalTitle{Optics Letters}} \textbf{51}, 1812 (2026).

\bibitem{Paschotta2004} R. Paschotta, \enquote{Noise of mode-locked lasers (Part I): numerical model,} {\protect\JournalTitle{Appl. Phys. B}} \textbf{79}, 153–162 (2004).


\bibitem{Agraval} G.~P. Agrawal, \textit{Nonlinear Fiber Optics}, 5th edition (Academic Press, 2013).

\bibitem{Lindberg} 
R.~Lindberg, P.~Zeil, M.~Malstrom, F.~Laurell, and V.~Pasiskevicius, \enquote{Accurate modelling of high-repetition rate ultrashort pulse amplification in optical fibers,}  {\protect\JournalTitle{Sci. Rep.}} \textbf{6}, 34742 (2016).

\bibitem{DuttaGupta2022} B. Dutta Gupta, I. Hendry, S. Tang, S. Coen, M. Erkintalo, and S. G. Murdoch, \enquote{Modeling of mode-locked fiber lasers and amplifiers with accurate saturable gain evaluation,} {\protect\JournalTitle{J. Opt. Soc. Am. B}} \textbf{39}, 3223–3230 (2022).

\bibitem{Hult}  J. Hult, \enquote{ A fourth-order Runge - Kutta in the interaction picture method for simulating supercontinuum generation in optical fibers. } {\protect\JournalTitle{Journal of Lightwave Technology}} \textbf{25}, 3770-3775 (2007).

\bibitem{olson} N. A. Olsson \enquote{Lightwave Systems with Optical Amplifiers} {\protect\JournalTitle{Journal of Lightwave Technology}} \textbf{7}, 1071 (1989).

\bibitem{Beker} P.~C. Becker, N.~A. Olsson, and J.~R. Simpson, \textit{Erbium-Doped Fiber Amplifiers, Fundamentals and Technology}, (Academic Press, 1999).

\bibitem{hui} R. Hui, and M. O'Sullivan, \textit{Fiber Optic Measurement Techniques}, (Elsevier, 2009).

\bibitem{desert} E. Desurvire, \textit{Erbium-Doped Fiber Amplifiers: Principles and Applications}, ( John Wiley and Sons. Inc., 2002).

\bibitem{Novak}  S. Novak and A. Moesle, \enquote{ Analytic Model for Gain Modulation in EDFAs. } {\protect\JournalTitle{Journal of Lightwave Technology}} \textbf{20}, 975-985 (2002).


\bibitem{Jopson}	
R.~M. Jopson and A.~A.~M. Saleh
\enquote{Modeling of gain and noise in erbium-doped fiber amplifiers,}  \textit{ in Fiber Lasers Sources and Amplifiers, III,} {\protect\JournalTitle{M.~J.~F. Digonnet and E. Snitzer, Eds., Proc. SPIE}} \textbf{1581}, 114--119 (1992).

\bibitem{Marcuse} D. Marcuse, \textit{Theory of Dielectric Optical Waveguides}, 2nd edition (Academic Press, 1991).

\bibitem{Kozlov} M. Kozlov, M. Ivanov, C. J. McKinstrie, and P. Sidorenko, \enquote{Generalized method of moments for
investigation of ultra-fast fiber amplifiers with gain-managed nonlinearity,} {\protect\JournalTitle{Optics Express}} \textbf{33}, 6171 (2025).


\bibitem{Mayer} A. S. Mayer, W. Grosinger, J. Fellinger,
G. Winkler, L. W. Perner, S. Droste, S. H. Salman, C. Li, C. M. Heyl, I.
Hartl, and A. H. Heckl, \enquote{Flexible all-PM NALM Yb:fiber laser design for frequency comb applications: operation regimes and their noise properties, } {\protect\JournalTitle{Optics Express}} \textbf{28}, 18946 (2020).

\bibitem{Canella} F. Canella, on behalf of L.M. Molteni, S. Cialdi, P. Laporta,
N. Coluccelli and G. Galzerano, \enquote{Realization and characterization of a low intensity noise ultrafast Yb-doped fiber amplifier, } {\protect\JournalTitle{IL NUOVO CIMENTO}} \textbf{44 C}, 23 (2021).

\end{thebibliography}
\end{document}